\documentclass[preprint,compress,3p]{elsarticle}
\usepackage{amsmath}
\usepackage{amsfonts}
\usepackage{amssymb}
\usepackage{float}
\usepackage{subfig}

\usepackage{cleveref}

\usepackage[usenames,dvipsnames]{color,colortbl}

\usepackage{graphicx}
\usepackage[export]{adjustbox}[2011/08/13]

\begin{document}
\begin{frontmatter}
\journal{Biomechanics and Modeling in Mechanobiology}
\title{Stresses and fluid flow in lamina cribrosa through anisotropic poroelasticity}
\author[1,2]{Riccardo Cavuoto}
\author[3]{Sofia Damian}
\author[3]{Luca Deseri\corref{cor}}\ead{ luca.deseri@unitn.it}
\author[1]{Massimiliano Fraldi\corref{cor}}\ead{fraldi@unina.it}
\author[8]{Brent Siesky}
\author[8]{Alice Verticchio}
\author[8]{Alon Harris}
\author[7]{Giovanna Guidoboni}

\address[1]{Department of Structures for Engineering and Architecture, University of Naples "Federico II", via Claudio 21, 80125 Naples, Italy \\ email: riccardo.cavuoto@unina.it, fraldi@unina.it}
\address[2]{Department of Neuroscience, Reproductive Sciences and Dentistry, University of Naples "Federico II", via Pansini 5, 80131 Naples, Italy}
\address[3]{Department of Civil, Environmental and Mechanical Engineering, University of Trento, via Mesiano 77, 38123 Trento, Italy \\ email: sofia.damian@unitn.it, luca.deseri@unitn.it}
\address[7]{Maine College of Engineering and Computing, University of Maine, Orono, ME, USA \\ email: giovanna.guidoboni@maine.edu}
\address[8]{Department of Ophthalmology, Icahn School of Medicine at Mount Sinai, New York, NY, USA \\ email: alice.verticchio@mssm.edu, brent.siesky@mssm.edu, alon.harris@mssm.edu}
\cortext[cor]{Corresponding authors}

\begin{abstract}
\indent
To investigate the mechanical correlations between intraocular pressure (IOP) variations and glaucoma, this study presents a linear transversely isotropic poroelastic model of the lamina cribrosa (LC) based on Reissner–Mindlin plate theory. A key feature of the proposed framework is its analytical tractability, which allows the governing poroelastic equations to be solved in closed form under appropriate mechanical and hydraulic boundary conditions. Within this setting, linearity is used to capture the reversible component of the tissue response, providing a baseline description of the coupled solid–fluid feedback on which more complex time-dependent phenomena, such as viscoelastic effects and remodelling, may build.
The results indicate that both strain and stress measures (in the form of shear strain and deviatoric stress measures) peak in the peripheral region of the LC, which is currently suspected to be the initial site of glaucomatous damage.
These quantities increase with IOP, suggesting a pressure-dependent mechanical insult to the retinal ganglion cell (RGC) axons.\\
\indent In parallel, the model predicts a monotonic reduction in fluid content as IOP rises, which may contribute to ischemic phenomena and disc haemorrhages.
The influence of material anisotropy was also examined, revealing that isotropic assumptions tend to overestimate the fluid content while underestimating shear strain.
Given the current experimental challenges in measuring blood flow within the LC, the proposed model provides a valuable framework for exploring the coupled mechanical–hemodynamic behavior of the tissue and for inverse estimation of its mechanical parameters, such as the stiffness of the opening for the central retinal vessels.

\end{abstract}

\end{frontmatter}
\section{Introduction}
\label{intro}
Glaucoma is a chronic eye pathology and one of the major causes of blindness worldwide \cite{Foster2005,Flaxman2017}. This is characterized by optic nerve head (ONH) damage, which are associated with the loss of the retinal ganglion cell (RGC) axon bundles, responsible for the delivery of the information collected from the retina to the brain \cite{Quigley1995,Weber1998}. Damage to the RGCs \cite[and references cited therein]{Fechtner1994,Burgoyne2011} can be correlated to (i) biomechanical insults due to the intraocular pressure (IOP) increase (biomechanical hypothesis), and/or (ii) malnourishment given by blood flow impediment (vascular hypothesis). The major site of RGCs axon bundles damage is the Lamina Cribrosa (LC) \cite{Bellezza2003,Anderson1974}, a sieve-like structure made of collagen beams, blood, and interstitial flow, where RGCs axons are collected to form the ONH. On the one hand, the LC acts as a structural protection of the neural components against mechanical insults. On the other hand, this structure is subject to a mechanical pressure load, also referred to as translaminar pressure difference, given by the difference among the IOP inside the eye, and the retrolaminar tissue pressure (RLTp) within the optic nerve. When the IOP is out of its physiological range, the translaminar pressure difference gives rise to LC deformations and strains that may lead to important RGCs insults. To understand the behavior of this tissue and correlate it to glaucomatous damage, many clinical studies have been performed on animal and human LC, both in-vivo \cite{Beotra2018,Midgett2019,Czerpak2024,Hannay2024} and ex-vivo \cite{Midgett2017}. To the same extent, also mathematical and computational models of the whole eye, of the posterior pole, or of the sole lamina have been provided during the years \cite{Dongqi1999,Edwards2001,Voorhees2017,Sander2006,Sigal2004}. Some examples of mathematical models of the LC can be found in the works of Dongqi et al. \cite{Dongqi1999} and Edwards et al. \cite{Edwards2001}. In the first one, the lamina is modelled as a thin, isotropic elastic plate, and some important mechanical features as the retrolaminar displacement under an applied pressure, have been investigated by assuming a small displacement regime. In the second case, the Authors computed stresses and strains in the LC by modelling it as a thin plate undergoing large deformations, and considering the lamina as a homogeneous and isotropic solid. With regard to computational models, many different approaches have been proposed during the years. An example is the work of Voorhees et al. \cite{Voorhees2017}, where a two-scale approach has been used. In particular, the Authors developed a mesoscale model of the ONH starting from imaging data and then, by using the outcomes of this model as boundary conditions, they defined a 3D microscale model of the LC. Another kind of computational model has been proposed by Sander et al. \cite{Sander2006}, where the microstructure of the lamina has been taken into account through cellular solid model of the laminar plates, in particular considering a unit cell derived from the repetition of octagonal plates. Finite element models of the LC can be found in the work of Sigal et al. \cite{Sigal2004}, where the detailed modelling of the human ONH and computational techniques are used to quantify stresses and strains induced by the IOP. Full 3D eye-specific finite element models of the human eye posterior pole including both the LC beams and the interspersed neural tissue can be found in the works of Karimi et al. \cite{Karimi2021,Karimi2022}.
Many of these studies focused on retrieving stresses and strains in the solid components of the LC (i.e. collagen beams), as these are supposed to play an important role in the mechanical interactions and RGCs insults. However, this may a priori neglect the influence of the vascular hypothesis on the glaucomatous damage. To account for the possible neural death due to malnourishment, indeed, blood flows \cite{Geijer1979} should also be taken into account when modelling the tissue. A mechanical model that does so naturally is poroelasticity \cite{causin2014,Ayyalasomayajula2016,Tatone2019}. 
Other two important aspects that should be considered to foster novel understandings on the biomechanics of the lamina are the anisotropy of the medium \cite{Roberts2009} and the effective aspect ratio of the lamina, which suggest that the LC is better approximated by a thick plate model than by a thin one, and that the passage of the axons gives a preferential plane of isotropy. 
In this respect, then, in the present work a transversely isotropic linear poroelastic model of the LC, based on the Reissner-Mindlin theory of plates \cite{Reissner1945,Mindlin1951} is put forward to study the effects of variations of the intraocular pressure IOP on the tissue biomechanical response. \\ 
The goal of the model is to look beyond the complex and rich mechanical behaviour of the lamina and obtain insights on driving reversible mechanisms that lie at the foundation of the response of this particular tissue. 
In fact, as is widely documented in the literature \cite{Karimi2021,causin2014,Voorhees2017,Li2020}, the lamina cribrosa exhibits a complex mechanical behaviour, including anisotropy, nonlinearity, viscoelasticity, and, over longer time scales, adaptive processes such as growth and remodelling. 
With the aim of isolating, within a simplified and analytically tractable setting, the main elastic and poroelastic mechanisms that may underlie its response, unveiling some still unexplored aspects at the basis of the mechanobiology of the lamina, other mechanical effects have thus been intentionally neglected at this stage.
In this respect, the adoption of a linear elastic framework should not be interpreted as assuming that the lamina cribrosa is intrinsically linear in all circumstances. Rather, it reflects the
view that, under operating conditions, i.e. physiological and pathological ones, the elastic part of the response may reasonably be treated as a first-order approximation. Indeed, although nonlinear responses can certainly be observed in ex vivo tests, it is more difficult to regard the in-vivo behavior of the lamina cribrosa as involving large purely elastic strains, given the strong geometric and mechanical compatibility constraints imposed by the surrounding tissues and by the neural bundles crossing the structure.
Excessively large reversible elastic deformations would imply severe local kinematic changes in a highly constrained biological environment, which would hardly be compatible with the preservation of the local tissue architecture and of the neural pathways traversing the lamina. 
This suggests that the large deformations reported in physiological or pathological settings are likely to reflect, at least in part, the combined action of additional mechanisms such as viscoelasticity, fluid effects, and remodeling, rather than purely nonlinear elasticity alone. 
It is in fact reasonable that some deviated geometrical configurations  observed experimentally in the LC might not be entirely reversible and thus the actual elastic deformation in response to loading conditions could still be assumed as linear.
Within this perspective, the value of the present linear model lies precisely in its analytical tractability. 
The possibility of deriving closed-form solutions makes it possible to identify the role of the governing mechanical parameters, to interpret the coupling between deformation and fluid-content variation, and to discuss qualitative trends in a transparent way. 
This level of control would be much harder to retain in a substantially more detailed constitutive framework. 
The intention of the present model is therefore to capture the elastic mechanisms that may act as driving factors for more complex and time-dependent phenomena, rather than to reproduce such phenomena directly.\\
The model is developed starting from the theory of poroelasticity \cite{BIOT1941,Wang2000,Alaimo2019FractionalPoroelasticity,COUSSY2004}. Equilibrium equations describing the problem of the Lamina Cribrosa have then been retrieved. Once the ansatz associated to the Reissner-Mindlin plate, and the suitable mechanical and hydraulic boundary conditions have been applied, it was possible to derive the analytical solutions governing the problem. In so doing, it was possible to analyze the hemodynamic component of the LC, thus retrieving the fluid content behavior when the IOP approaches both the hypotension and hypertension regimes. 
At the same time, we derived both a deviatoric stress measure and the shear strain, which are markers of possible RGCs axonal injuries. The numerical outcomes suggest high levels of both these elastic measures in the peripheral zone of the LC, typically the site of the first glaucomatous damage, where the RGCs responsible of the peripheral visual field cross the lamina. 
This trend increases with higher levels of IOP, thus suggesting the possible pressure-dependence of the mechanical insult to the RGC axon bundles. In terms of the fluid content, we observed a monotonic decrease of the blood flow as the IOP increases, which may lead to some indirect and qualitative conclusions in terms of ischemic behavior and disc hemorrhages. 

More detailed information could be obtained by explicitly including quantities such as the blood flow velocity and perfusion rate description in the analytical approach \cite{Tatone2019,Wang2017}. Yet, since this lies beyond the purpose of this work, the decrease in the fluid content can only indicate a local depletion of the effective fluid phase within the tissue and may therefore suggest a reduction in the availability of nutrients and oxygen, consistent with an increased risk of ischemic insult. 
The present formulation is developed within a homogenized poroelastic setting, in which the fluid is treated as an effective macroscopic phase and the experimentally observed pressure-driven
transport is represented in the most natural continuum form, namely through Darcy’s law.
At a finer microscale level, more detailed transport descriptions may be introduced, including capillary-flow mechanisms, deformation-dependent permeability, and possibly additional interaction effects. 
However, once such microscopic details are incorporated explicitly, a model would require a substantially richer multi-scale description, potentially involving further mechanisms whose inclusion would go well beyond the scope of the present analytical study. 
The aim here is therefore not to provide an exhaustive microscale fluid model, but rather to retain the simplest homogenized description capable of representing the macroscopic coupling between deformation and fluid transport for gaining fundamental insights into how fluid flow, fluid content, interstitial pressure and stresses all interact to determine critical over-loaded regions as well as anomalous distributions of nutrients
supply.
Having said this, it is also worth noting that, currently, blood flow measurements inside the LC tissue are quite difficult to obtain.

In this fashion, the model here proposed may give the opportunity of analyzing the fluid behavior as well as combining the mechanical and hemodynamic effects that may be present in the Lamina Cribrosa. 
Some investigations have been performed on the effect of the anisotropy in the model, highlighting how the isotropic assumption may lead to overestimation of the fluid content and underestimation of the shear strain. 
In addition, the model can be used to perform inverse analyses to retrieve specific LC's mechanical parameters, for which there are few data in the literature because of measurements difficulties. Among these, there is the stiffness of the opening for central canal retinal vessels. To this extent, here we have conducted some parametric analysis by considering various stiffness values, showing that this parameter provides a non-negligible effect on stresses and deformations of the LC tissue. \\ 
\indent The paper is organized as follows. In Sect. 2 the poroelastic theory is presented; in Sect. 3 poroelasticity is specialized to the case of the Lamina Cribrosa and the model is introduced, the results are also reported; finally, in Sect. 4 we discuss the outcomes.

\section{Governing equations of poroelasticity}
Poroelasticity \cite{BIOT1941,Wang2000,COUSSY2004} represents a well-established theoretical framework to describe the response of many (hard and soft) biological tissues \cite{COWIN2007,FRALDI2018,CAROTENUTO2021}. The general equations of poroelasticity for saturated solids, i.e. with completely fluid-filled cavities, indeed involve homogenized properties that mediate the matrix and the fluid mechanical properties through microscopic information about the extent of reference porosity and, in general, its spatial distribution. Recalling, in what follows, the general equations of linear poroelasticity, the first basic hypothesis considers the Terzaghi decomposition of the effective stress $\boldsymbol{\sigma}^{eff}$ into the sum of the solid stress $\boldsymbol{\sigma}$ and the fluid pressure $p-p_0$, where $p_0$ is a baseline pressure:

\begin{equation}\label{Terzaghi}
    \boldsymbol{\sigma}^{eff}=\mathbb{C}:\boldsymbol{\varepsilon}=\boldsymbol{\sigma}+\mathbf{A}(p-p_0)
\end{equation}
in which $\mathbb{C}$ is the fourth-order symmetric (major and minor symmetries) stiffness tensor of the \textit{drained} homogenized medium --i.e. of the porous medium in which fluid flow is permitted-- while, in absence of any inelastic contribution, the tensor $\boldsymbol{\varepsilon}=\text{sym}(\mathbf{u}\otimes\nabla)$ is the overall strain given by the symmetrized displacement gradient. The matrix $\mathbf{A}$ is commonly referred to as the Biot effective stress coefficient symmetric tensor, which measures the contribution of pore strain to the increment of fluid pressure. Its generic expression reads as \cite{COWIN2007}:

\begin{equation}
    \mathbf{A}=\left(\mathbb{I}-\mathbb{C}:\mathbb{S}^{(m)}\right):\mathbf{I} \; ,
\end{equation}
where $\mathbb{S}^{(m)}=\left[\mathbb{C}^{(m)}\right]^{-1}$ denotes the compliance tensor of the isolated solid medium. By solving for the solid stress in Eq.~\eqref{Terzaghi}, the balance equations in quasi-static conditions are written with respect to the solid frame:

\begin{eqnarray} \label{momb}
    &\boldsymbol{\sigma}=\mathbb{C}:\boldsymbol{\varepsilon}-\mathbf{A}(p-p_0) \; ,\notag \\
    & \nabla\cdot \boldsymbol{\sigma}=\textbf{0} \; , \quad \boldsymbol{\sigma}=\boldsymbol{\sigma}^{T} \; .
\end{eqnarray}

The presence of the fluid pressure as additional macroscopic field implies that the mechanical equilibrium problem is naturally coupled with the fluid conservation equation. In classical poroelasticity, the variation in fluid content $\zeta=\varphi-\varphi_{0}$ is conveniently introduced as fluid field variable, and the generic form of fluid mass balance can be written as

\begin{equation}\label{massb}
    \frac{\partial \zeta}{\partial t} + \nabla \cdot \mathbf{q}_F = \Gamma_F \; ,
\end{equation}
where the vector $\mathbf{q}_F$ represents the fluid flux, whereas $\Gamma_F$ is a source/sink term modelling potential accumulation or depletion of fluid. 
In this respect, thermodynamical considerations lead to introduce a constitutive equation connecting the variation in fluid content $\zeta$ in saturated media as the result of pore geometric strain and fluid pressure action, that is
\begin{equation}\label{Zeta}
    \zeta = \mathbf{A}:\boldsymbol{\varepsilon} + M^{-1}(p-p_0)=\mathbf{A}:\mathbb{S}:\boldsymbol{\sigma}+C^{eff}(p-p_0) \; , \notag\\
\end{equation}
where $\mathbb{S}$ represents the compliance tensor of the homogenized medium. For further developments, the equation above can be conveniently rewritten as a function of the pore pressure:
\begin{equation}
    p-p_0 = M(\zeta-\mathbf{A}:\boldsymbol{\varepsilon}) \; ,    
\end{equation}


\noindent where the coefficients $C^{eff}$ and $M^{-1}$ are denoted respectively as Biot's effective modulus and compressibility coefficient. {The balance equations reported above and the derivation proposed in what follows, are both well-known in the mechanics community, thus the interested reader can find more details about them in the several books and referecens on the topic, namely \cite{COWIN2007,CAROTENUTO2021,Cavuoto2026}. In particular, micromechanical considerations lead to derive the Biot modulus as 
\cite{COWIN2007,CAROTENUTO2021,Cavuoto2026}

\begin{equation}
    C^{eff}=(\mathbf{I}:\mathbb{S}:\mathbf{I})-(\mathbf{I}:\mathbb{S}^{(m)}:\mathbf{I}) +\varphi \left(K_F^{-1}-(\mathbf{I}:\mathbb{S}^{(m)}:\mathbf{I})\right) \; ,
\end{equation}


\noindent where $\varphi$ indicates the tangent porosity and $K_F$ the fluid bulk modulus. In most applications, the fluid can be considered as incompressible, and this is the case for blood. It is also useful to note that Biot moduli are related to each other through the following expression: 

\begin{equation}\label{Biot modulus}
    M^{-1}= C^{eff}-\mathbf{A}:\left(\mathbb{S}-\mathbb{S}^{(m)}\right):\mathbf{I}=C^{eff}(1-\mathbf{A}:\mathbf{B}) \; .
\end{equation}

Herein, the emerging poroelastic coupling coefficient $\mathbf{B}$ is the Skempton coefficient tensor:

\begin{equation}\label{Skempton}
    \mathbf{B}=\frac{1}{C^{eff}}\left[\mathbf{I}:\left(\mathbb{S}-\mathbb{S}^{(m)}\right)\right]=\frac{1}{C^{eff}}\mathbf{A}:\mathbb{S} \; .
\end{equation}


This tensor represents the linear connection between the solid stress and the fluid pressure when the medium is tested in  \textit{undrained} conditions, i.e. when the fluid flow is impeded and $\zeta = 0$. By imposing this condition in Eq. \eqref{Zeta} the following expression involving both the hydraulic pressure, $p^{(u)}$, and the stress, $\boldsymbol{\sigma}^{(u)}$, evaluated in perfectly undrained condition are obtained:

\begin{equation}
    p^{(u)}-p_0= - \frac{1}{C^{eff}} \mathbf{A}:\mathbb{S}:\boldsymbol{\sigma}^{(u)}=- \mathbf{B}:\boldsymbol{\sigma}^{(u)} \; .
\end{equation}

The estimation of the strain $\boldsymbol{\varepsilon}^{(u)}$ in undrained conditions through the use of Eq. \eqref{Terzaghi} leads to eventually derive an expression for the undrained elastic constants 
\begin{eqnarray}\label{undrained}
&\boldsymbol{\varepsilon}^{(u)}=\mathbb{S}^{(u)}:\boldsymbol{\sigma}^{(u)} \; ,\notag\\    
    &\displaystyle\mathbb{S}^{(u)}=\mathbb{S}:\left[\mathbb{I}-(\mathbf{A}\otimes\mathbf{B})\right]=\mathbb{S}-\frac{1}{C^{eff}}\left[\mathbf{I}:\left(\mathbb{S}-\mathbb{S}^{(m)}\right)\right]\otimes \left[\left(\mathbb{S}-\mathbb{S}^{(m)}\right):\mathbf{I}\right] \; .
\end{eqnarray}

To close the system of equations given by Eqs. (\ref{momb}) and (\ref{massb}) in the two unknown fields $(\mathbf{u},p)$ a relationship between fluid flow $\mathbf{q}_F$ and pressure $p$ is necessary. The required relation governing the fluid filtration is a constitutive equation that has its simplest form in the standard Darcy's Law. This relates the fluid flux to the pressure gradient linearly, i.e.:
\begin{equation} \label{darcy}
    \mathbf{q}_F=-\mu^{-1}\mathbf{k}\nabla p \; ,
\end{equation}
where $\mu$ is the fluid viscosity, $\mathbf{k}$ is the intrinsic permeability symmetric tensor and, lastly, $\nabla p$ is the pressure gradient.

\paragraph{Reduction to the isotropic case}

If isotropy of the homogenized and matrix constants is assumed, the above introduced relations can be simplified and the most of coupling poroelastic parameters reduce to scalar coefficients. In particular, the assumption of $\mathbf{A}=\alpha \mathbf{I}$ leads to the following constitutive relations:

\begin{eqnarray}\label{Isotropy}
    &\boldsymbol{\sigma}=\mathbb{C}:\boldsymbol{\varepsilon}-\alpha(p-p_0)\mathbf{I} \; ,\notag\\
    &\zeta = \alpha e + M^{-1}(p-p_0) \; ,
\end{eqnarray}
with $e=tr{\boldsymbol{\varepsilon}}=\nabla\cdot \mathbf{u}$ being the volumetric strain, while $\alpha= (1-K/K^{(m)})$ 
is typically called isotropic effective stress coefficient. In these conditions, equations are usually rewritten in terms of drained and undrained coefficients, directly measurable from the fluid-filled solid. To this aim, let us write the Skempton coefficient from Eq. \eqref{Skempton} under isotropy as a function of the drained Young modulus $E$ and Poisson ratio $\nu$:

\begin{equation}\label{isoSkemp}
    \mathbf{B}=\frac{B}{3}\mathbf{I}= \frac{\alpha}{C^{eff}}\left(\frac{1-2\nu}{E}\right)\mathbf{I} \; .
\end{equation}
\indent Relation \eqref{undrained} now reads as follows:
\begin{equation}
   \left[\frac{1+\nu^{(u)}}{E^{(u)}}\right] \mathbb{I}-\left[\frac{\nu^{(u)}}{E^{(u)}}\right](\mathbf{I}\otimes\mathbf{I})=\left[\frac{1+\nu}{E}\right] \mathbb{I}-\left[\frac{\nu}{E}+\frac{\alpha B}{3}\frac{1-2\nu}{E}\right](\mathbf{I}\otimes\mathbf{I})
\end{equation}
which, in combination with Eqs. \eqref{isoSkemp} and \eqref{Biot modulus}, leads to the following relations

\begin{eqnarray}
    &\displaystyle\alpha= \frac{3}{B}\frac{\nu^{(u)}-\nu}{(1-2\nu)(1+\nu^{(u)})} \; ,\\
    &\displaystyle C^{eff}=\frac{9}{E B^2}\left(\frac{\nu^{(u)}-\nu}{1+\nu^{(u)}}\right) \; ,\\
    &\displaystyle M^{-1}=\frac{9}{B^2}\frac{(\nu^{(u)}-\nu)(1+\nu)(1-2\nu^{(u)})}{(1-2\nu)(1+\nu^{(u)})^2} \; .
\end{eqnarray}

In the limit case of an incompressible isotropic matrix (i.e. $\nu^{(m)}\to 1/2^-$) and incompressible fluid, it is straightforward to verify that $\alpha=1$, while $C^{eff}=1/K$ and $B=1$. Furthermore, the undrained Poisson coefficient $\nu^{(u)}$ also takes the value of 1/2 as the compressibility factor $M^{-1}\to 0$. 

Under the assumptions of an isotropic pore distribution $\mathbf{k}=k \mathbf{I}$, the constitutive law governing filtration Eq. (\ref{darcy}) specializes in
\begin{equation}
    \mathbf{q}_F=-\mu^{-1} k\nabla p \; .
\end{equation}

These equations for the isotropic case will be useful for the comparison with the results retrieved from the anisotropic model of the Lamina Cribrosa.

\section{Influence of lamina cribrosa structure}

 The lamina cribrosa (LC) is a porous, sieve-like structure located at the back of the eye, at the level of the optic nerve head. 
It appears as a soft, disk-shaped tissue containing a series of small openings through which the axons of retinal ganglion cells pass to form the optic nerve. 
Hence, the LC plays a crucial role in supporting and protecting these axons and in preserving their structural and functional integrity, as they are responsible for the transmission of visual stimuli to the brain. 
The lamina cribrosa is therefore essential for the health of the optic nerve, and its maladaptive remodelling is potentially involved in several ocular diseases. 
Indeed, morphological and mechanical alterations within the LC are strongly correlated with altered intraocular pressure (IOP) and other primary factors involved in optic nerve damage and in the onset and progression of chronic glaucoma. 
The monitoring of structural changes in the optic nerve head (ONH), where the LC resides, is one of the most important clinical tools for assessing glaucoma progression.
In this respect, relevant indicators are related to LC deformation or clinical glaucomatous \lq\lq cupping\rq\rq, mainly expressed in terms of LC thickening and LC posterior displacement and excavation, which affect the physiological lamina cribrosa curvature index (LCCI) and may compromise ONH conditions. 
Besides geometrical factors, material remodelling may also occur, together with an increase in the connective tissue component, alterations of the laminar beam architecture, and a decrease in pore size. 
However, given the diagnostic relevance of morphological aspects, it is worth emphasising that the health of the ONH bundles also critically depends on the transport of oxygen and nutrients from the laminar capillaries through the laminar extracellular matrix, made of collagen beams, and finally to the peripheral and central axons of each bundle.}

 Therefore, the LC is well known to exhibit a complex mechanical behaviour, characterised by nonlinearity, anisotropy, viscoelasticity, and adaptive mechanisms \cite{Voorhees2017,Li2020}.
As for many other biological tissues, a fully detailed representation of its behaviour would require a high level of mechanical complexity, which may reduce the interpretability of the model and make it more difficult to isolate the role of the individual mechanisms involved. 
For this reason, the purpose of the present work is to introduce a simplified and analytically tractable model capable of capturing the main qualitative features of the LC and of delivering a closed-form solution for the interpretation of the key reversible aspects that drive time-dependent and remodelling mechanisms. 
Within this perspective, a linear poroelastic model, able to analyse the mutual interplay among deformation, stresses, stress gradients, and the transport of fluid carrying nutrients, can provide crucial insight into how LC morphological changes and altered micro-environmental hydraulic conditions cooperate in determining adverse adaptation. 
The use of a linear elastic contribution does not imply that the overall response of the lamina cribrosa is itself small or entirely reversible. Rather, in such a highly organised and constrained biological tissue, it is unlikely that the full deformation developing under physiological or pathological conditions can be entirely attributed to elasticity alone. A significant part of the response is more plausibly governed by anelastic, time-dependent, and adaptive mechanisms. Accordingly, the elastic contribution may be interpreted as the reversible part of a broader mechanical behaviour, and thus as a first-order component that can reasonably be linearised. Moreover, as will be shown for the geometries and loading conditions considered here, the resulting elastic strains remain sufficiently small to be consistent with the assumptions of linear theory. On this basis, a linear poroelastic formulation is adopted here as a tractable framework for investigating the coupled interplay between deformation, stress, and fluid transport.

\paragraph{LC micro-structural implications}  From a mechanical standpoint, the particular porous architecture of the LC consists of a matrix populated by cylindrical cavities developing through the thickness direction, within which the softer axons pass. This defines a first type of porosity (\textit{primary porosity}, $\phi$, henceforth), which enters the homogenization procedure, affects the elastic moduli of the LC, and contributes to the anisotropic character of the homogenized mechanical response. At a lower scale, the matrix itself exhibits a permeable fluid-saturated structure, with a solid component containing collagen fibrils mainly oriented in the cross-sectional plane of the LC. The presence of cylindrical cavities together with this fibrillar organization suggests describing the effective response of the lamina cribrosa as that of a transversely isotropic poroelastic medium. It is worth noting that the choice of this symmetry class directly follows from the classical interpretation of the underlying microstructure within homogenization theory. Indeed, since the collagen fibrils exhibit no preferred direction in the cross-sectional plane, at least in a statistical or averaged sense, this plane may be regarded as a plane of isotropy. By contrast, the structurally distinct out-of-plane direction naturally identifies an axis of material symmetry, thus leading to a transversely isotropic response at the effective continuum level. A second type of porosity, namely the mesoscale intrinsic porosity of the solid matrix (\textit{secondary porosity}, $\varphi$, in the sequel), is instead associated with the permeability of the poroelastic medium and therefore enters the Darcy-type constitutive equations governing the fluid flux. This porosity combines with capillary orientation, which is essentially radial, to determine the effective permeability of the system, which in turn steers fluid pathways within the laminar extracellular space. This permeability may also be regarded as transversely isotropic. Therefore, while the matrix material may be considered intrinsically isotropic, the effective poroelastic response is modelled as transversely isotropic. Denoting the transverse and axial components by the subscripts $t$ and $z$, respectively, the homogenized compliance tensor is determined as follows:
\begin{equation}\label{Stiso}
    \mathbb{S}=\left(
\begin{array}{cccccc}
 \frac{1}{E_t} & -\frac{\nu _t}{E_t} & -\frac{\nu
   _z}{E_z} & 0 & 0 & 0 \\
 -\frac{\nu _t}{E_t} & \frac{1}{E_t} & -\frac{\nu
   _z}{E_z} & 0 & 0 & 0 \\
 -\frac{\nu _z}{E_z} & -\frac{\nu _z}{E_z} &
   \frac{1}{E_z} & 0 & 0 & 0 \\
 0 & 0 & 0 & \frac{1}{2 \mu _z} & 0 & 0 \\
 0 & 0 & 0 & 0 & \frac{1}{2 \mu _z} & 0 \\
 0 & 0 & 0 & 0 & 0 & \frac{\nu _t+1}{E_t} \\
\end{array}
\right) \; .
\end{equation}

Furthermore, by setting

\begin{equation}\label{diag}
    \mathbf{A}=\text{Diag}\{\alpha_t,\alpha_t,\alpha_z\} \ ,
\end{equation}



\noindent  and defining the Young's modulus of the isotropic matrix as $E_m$, the homogenized Young's modulus in the transverse plane as $E_t$ and that in the axial direction as $E_z$, one explicitly has

\begin{eqnarray}\label{Atiso}
\alpha_t = 1-\frac{E_t \left(1-2 \nu _m\right) \left(1+\nu
   _z\right)}{E_m \left(1-\nu_t -2 \nu_z^2 E_t/E_z \right)}, \quad 
   & \alpha_z& = 1-\frac{E_z}{E_m}\frac{\left(1-2 \nu _m\right) \left(1-\nu
   _t+2 \nu _z E_t/E_z \right)}{1-\nu _t -2 \nu _z^2 E_t/E_z}
\end{eqnarray}

\noindent and, through the use of Eqs. \eqref{Biot modulus} and \eqref{Skempton}, the Biot compressibility modulus becomes



\begin{equation}\label{Mtiso}
M^{-1}=\frac{1}{K_m}\frac{(1-\nu_t-2\nu_z^2 E_t/E_z)(1-\varphi)-(1-2\nu_m)\left[1-\nu_t+2(1+2\nu_z)E_t/E_z \right](E_z/E_m)/3}{1-\nu _t - 2 \nu_z^2 E_t/E_z} \; ,
\end{equation}
where $K_m$ is the bulk modulus of the matrix.\\
\indent In a transversely isotropic framework Darcy's Law (\ref{darcy}) assumes a simple form given that the permeability tensor has a diagonal form of the kind
\begin{equation} \label{tidarcy}
    \mathbf{K}=\text{Diag}\{k_t,k_t,k_z\} \; ,
\end{equation}
where $k_t$ and $k_z$ are the permeabilities in the isotropic plane and in the transversal direction, respectively. \\
 The present formulation is developed within a homogenised poroelastic setting, in which the fluid is treated as an effective macroscopic phase and the experimentally observed pressure-driven transport is represented in the most natural continuum form, namely through Darcy’s law. At a finer microscale level, more detailed transport descriptions may be introduced, including capillary-flow mechanisms, deformation-dependent permeability, and possibly additional interaction effects \cite{Alaimo2019FractionalPoroelasticity,FRALDI2018}. However, explicitly incorporating such microscopic details would require a substantially richer multiscale description, potentially involving further mechanisms whose inclusion would go well beyond the scope of the present analytical study. Therefore, rather than aiming to provide an exhaustive microscale fluid model, the present work retains the simplest homogenised description capable of representing the macroscopic coupling between deformation and fluid transport, thereby offering fundamental insight into how fluid flow, fluid content, interstitial pressure, and stresses interact in determining critically overloaded regions as well as anomalous distributions of nutrient supply.
\subsection{Governing equations for the LC problem}

\begin{figure}[!h]
    \centering
    \includegraphics[width=0.8\linewidth]{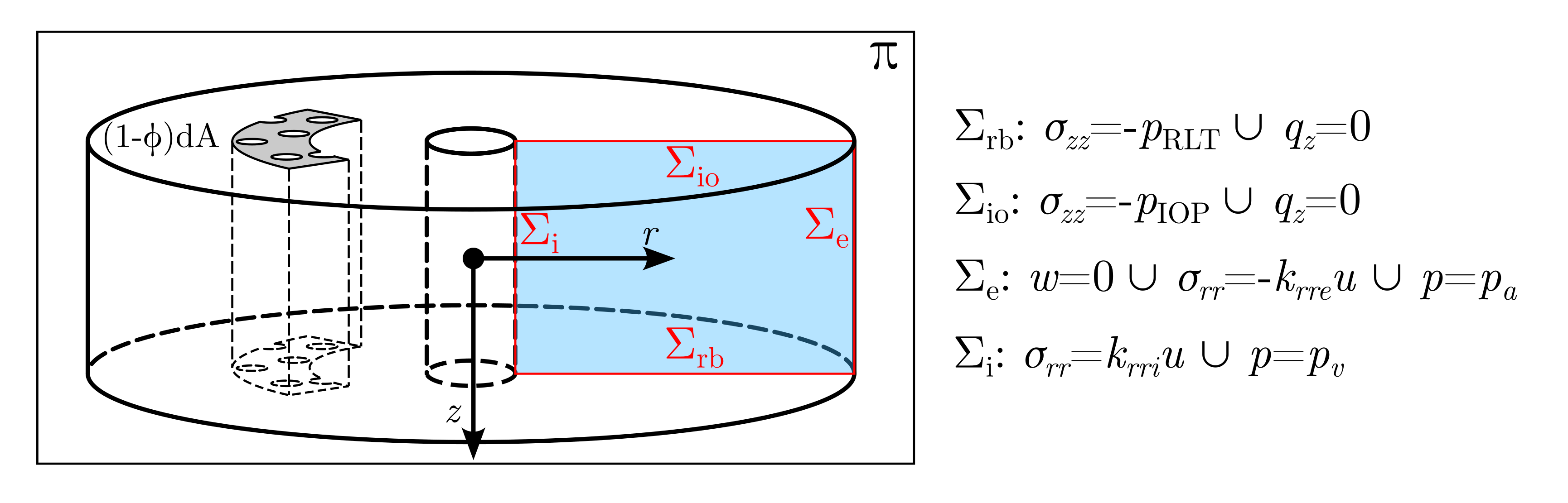}
    \caption{Sketch of the boundary conditions applied to the proposed model.}
    \label{fig:bc}
\end{figure}

The disk-shaped LC allows for a series of observations that enable simplified solutions to the system of differential equations governing momentum and fluid mass balance for a transversely isotropic material. In particular, the cylindrical geometry permits the exploitation of the problem’s circumferential symmetry, implying that the solution, defined through the unknown fields $(\boldsymbol{u}, p)$, depends only on the radial and vertical coordinates. The two displacement field components, $\boldsymbol{u} = \{u, w\}$, represent the radial and vertical displacements, respectively, consistent with the assumed circumferential symmetry.\\
\indent Under these conditions the scalar form of the governing equations (\ref{momb}-\ref{massb}) for the transversely isotropic Lamina cribrosa specialize to
\begin{equation} \label{eq:ti_balancemass}
    k_t\frac{\partial^2 p}{\partial r^2}+k_z\frac{\partial ^2p}{\partial z^2}+\frac{k_t}{r}\frac{\partial p}{\partial r}=0 
\end{equation}
for the continuity equation, and 
 
\begin{equation} \label{eq:ti_momentumr}
    \frac{\partial^2 u}{\partial r^2}+\frac{E_t}{4\mu_z}\frac{\partial^2 u}{\partial z^2}+\frac{1}{r}\frac{\partial u}{\partial r}-\frac{u}{r}+\left(\frac{E_t}{4\mu_z}-\nu_z \frac{E_t}{E_z} \right)\frac{\partial^2 w}{\partial z\partial r}-\alpha_tE_t \frac{\partial p}{\partial r}=0\ \ ,
\end{equation}
\begin{equation}\label{eq:ti_momentumz}
    \frac{E_z-4\mu_z\nu_z}{E_z}\left(\frac{\partial ^2u}{\partial r\partial z}+\frac{1}{r}\frac{\partial u}{\partial z} \right)+\frac{\partial ^2w}{\partial r^2}+\frac{4\mu_z}{E_z}\frac{\partial ^2w}{\partial z^2}+\frac{1}{r}\frac{\partial w}{\partial r}-4\alpha_z\mu_z\frac{\partial p}{\partial z}=0
\end{equation}
for the momentum balance equations in the radial direction and vertical direction, respectively. 
\paragraph{Boundary conditions} The boundary conditions required are reported in Figure \ref{fig:bc}. The retrobulbar ($\Sigma_{rb}$) and intraocular ($\Sigma_{io}$) edges of the lamina cribrosa are loaded by the trans corneal pressure difference, IOP (above - $p_{IOP}$), and the RLT (below - $p_{RLT}$) pressures. To explore the effects of pathological conditions, the intraocular pressure is assumed to vary in a range of values from 5 mmHg to 30 mmHg (physiological values around 12 to 15 mmHg) \cite{guidoboni2014,Yan1998,Okonkwo2026,Girkin2024,Shen2025,Jonas2013}.\\
The RLT pressure is fixed at 7 mmHg (physiological baseline value). This loading condition is uploaded in the model by applying tractions at the intraocular and retrobulbar edges, namely $\boldsymbol{\sigma}\boldsymbol{n}=-p_{IOP}\boldsymbol{n}$ at the intraocular side and $\boldsymbol{\sigma}\boldsymbol{n}=-p_{RLT}\boldsymbol{n}$ at the retrobulbar side. From the hydraulic point of view, the intraocular and retrobulbar boundaries of the LC are impermeable, hence it is assumed there that the flux be null, $\boldsymbol{q}\cdot \boldsymbol{n}=0$. \\
\indent The outer surface of the cylindrical LC ($\Sigma_e$) -- separating the LC from the peripapillary sclera -- enforces continuity of the displacements between the sclera and the LC. 
 Null vertical displacement of the middle surface $w=0$ is imposed in order to prevent rigid vertical translation of the reduced plate section at the outer edge, while still allowing for through-thickness deformations at the LC-sclera interface, whereas the radial displacements depend on the stiffness of the sclera. To account for these stresses exchanged in the radial direction, we assume $\sigma_{rr}=-k_{rr,e}u(R)$, with $u(R)$ being the radial displacement on the lateral surface of the LC, while $k_{rr,e}=E_s/h$ is defined as the ratio between the sclera Young's Modulus in the radial direction and its height.  It is worth noting that $k_{rr,e}$ represents a physically motivated simplified estimation of the elastic radial interaction between the LC and the surrounding peripapillary sclera. As far as the hydraulic boundary conditions are concerned, on this surface, the pressure must balance that of the arteriolar vessels that meet the LC, so $p(R)=p_a=30 $ mmHg.

The inner surface ($\Sigma_i$), representing the central cylindrical cavity hosting the central retinal artery and central retinal vein, follows similar boundary conditions to the external surface: $\sigma_{rr}=k_{rr,i}u(R_i)$ for the applied tractions and $p=p_v=20$ mmHg for the hydraulic one.
To solve the equilibrium problem for the transversely isotropic cylindrical LC a model for transversely isotropic poroelastic Mindlin plate is presented below, following a series of considerations on the type of solutions searched for. 

\paragraph{Mindlin poroelastic plate model} 
Dimensionally-reduced formulation hold many advantages in the understanding of the mechanics of slender structures. These have gained success over the years in modelling several applications \cite{Bernard2024}. In this section a dimensionally-reduced model is developed for the poroelastic LC that is based on Mindlin's plate theory. \\
\indent The geometric characteristics of the LC, in fact, with their relatively high diameter-to-height ratios ($2R/h > 5$) fall in the so-called range of moderately thick plates. Accordingly, the kinematics of the LC can be assumed to follow:
\begin{equation}
   u(r,z)=U(r)+z\ \psi(r) \; , \quad w(r,z)=W(r)+\varepsilon_0\ z+ \varepsilon_1\ z^2 \; , \quad p(r,z)=P(r) \; . \label{eq:ansatz}
\end{equation}
\indent  This is the specific kinematic ansatz of the reduced Mindlin model, where the retained terms are exact within the assumed plate kinematics. In the transverse direction, an enrichment is introduced to properly account for non-trivial variations in the transverse normal stress, which is non-uniform for equilibrium requirements due to the different pressures applied on the upper and lower faces of the LC. Furthermore, the deformability in the thickness direction allows a squeezing mechanism, which affects the in-plane fluid transport. It is worth noting that despite the kinematic enrichment performed by retaining higher-order terms in the through-thickness direction for the vertical displacement, eq. (\ref{eq:ansatz}) $w(r,z)$, does not alter the formal truncation order of the reduced equilibrium equations. With respect to the expression chosen for the pressure, this is adopted as part of the reduced homogenization formulation and is expected to remain valid within the regime considered here, namely in the presence of impermeable upper and lower boundaries and within the linear Mindlin-type approximation. Indeed, as the IOP and RLTp enter the model as external mechanical loads on the solid skeleton, and the two faces are assumed impermeable, the internal pore fluid is not placed in direct hydraulic communication with the exterior through the thickness, and the pore-pressure gradient is therefore substantially reduced.

\indent Hence, given (\ref{eq:ansatz}), the components of the strain tensor $\boldsymbol{\varepsilon}$, in cylindrical coordinates, read as follows:
\begin{table}[!h]
    \centering
    \begin{tabular}{l l l}
    $\varepsilon_{rr}=U'(r)+z\ \psi'(r)$ & $\varepsilon_{\theta \theta}= U(r)/r+z\ \psi(r)/r$ & $\varepsilon_{zz}=\varepsilon_0+2\varepsilon_1\ z$\\
    $\varepsilon_{\theta z}=0$ & $2\varepsilon_{rz}=\psi(r)+W'(r)$ & $\varepsilon_{r\theta}=0$ ,
    \end{tabular}
    \label{tab:strain}
\end{table}

\noindent where the symbol $'$ indicates derivative with respect to the variable $r$. Two major aspects that distinguish eq. (\ref{eq:ansatz}) from what typically found in the literature are: (i) a dependence of the vertical displacement on through-thickness coordinate is added to account for axial deformations in that direction -- a fact that ultimately leads to having all the three extensional deformations ($\varepsilon_{rr},\varepsilon_{\theta \theta},\varepsilon_{zz}$) linear in $z$; (ii) pressure $p$ being constant through the thickness is merely a consequence of the boundary conditions in the thickness direction where a null flux (and pressure gradient) is asked for in that direction \cite{sladek2015}.\\
\indent Following eqs. (\ref{momb}), balance of linear momentum in the radial and vertical direction, and fluid mass balance are obtained by integration along $z$:
 
\begin{eqnarray}
    &\frac{E_t (1-\nu_z^2 E_t/E_z)(-U+rU'+r^2U'')}{r(1+\nu_t)(1-\nu_t-2 \nu_z^2 E_t/E_z)}-\alpha_t\ r \ P'=0 \; ,\label{eq:mombr}\\
    &h\left(\frac{\mu_z(\psi(r)+W'(r))}{r}+\mu_z(\psi'(r)+W''(r)) \right)+q=0 \; ,\label{eq:mombz}\\
    &k_t \left(\frac{P'}{r}+P''\right)=0 \; .\label{eq:massbal}
\end{eqnarray}
where $q=p_{IOP}-p_{RLT}$.\\
\indent Additionally, due to the reduction to the middle plane, the balance of angular momentum about the $\theta$ direction must be invoked, thereby providing the fourth differential equation to match the number of unknown fields. This balance is obtained by multiplying the linear momentum equation in the $r$ direction by $z$, and then integrating through the thickness, yielding:
 
\begin{equation} \label{eq:angmombal}
    \left(\frac{(1-\nu_z^2 E_t/E_z)(1-\nu_t)}{(1-\nu_t-2 \nu_z^2 E_t/E_z)}\frac{E_t h^3}{12(1-\nu_t^2)}\right)(-\psi(r)+r\psi'(r)+r^2\psi''(r))=h\ r^2\mu_z(\psi(r)+W'(r)) \; . 
\end{equation}
\indent The system of four differential equations defined by (\ref{eq:mombr}-\ref{eq:angmombal}) can easily be decoupled and solved, obtaining a closed-form solution straightforwardly. 
Accordingly:
\begin{eqnarray}
    &U(r)=c_2\frac{R}{r}+c_3\ \frac{r}{R}-c_1\mathcal{A}\ \frac{r}{R}\left(1-2\log \frac{r}{R}\right)\\
    &P(r)=c_4+c_1 \log \frac{r}{R}\\
    &W(r)=\frac{3q\mathcal{A}}{4h^3\alpha_t}\frac{r^4}{R^4}+\left(\frac{c_6}{2}-\frac{c_7}{4}-\frac{q}{4h\mu_z}\right)\frac{r^2}{R^2}+c_8+\left(c_5+c_7\frac{r^2}{2R^2}\right)\log\frac{r}{R}\\
    &\psi(r)=-\frac{3q\mathcal{A}}{h^3\alpha_t}\frac{r^3}{R^3}-c_6\frac{r}{R}-\left(c_5+\frac{h^2\alpha_tc_7}{24\mathcal{A}\mu_z}\right)\frac{R}{r}+c_7\frac{r}{R}\log \frac{r}{R} \; ,
\end{eqnarray}
where 
 $\mathcal{A}=\frac{\alpha_t(1+\nu_t)(1-\nu_t-2 \nu_z^2 E_t/E_z)}{4E_t (1-\nu_z^2 E_t/E_z)}$, $R$ is the external radius of the cylindrical LC, and the $c_i$ are eight integration constants that are uniquely determined using the eight boundary conditions presented below. The natural boundary conditions for the plate model to uniquely identify these constants can be obtained starting from the boundary conditions discussed above, and read as follows:
\begin{align*}
    \int_{-h/2}^{h/2}p(R,z)\mathrm{d}z=hp_a &\; , \quad \int_{-h/2}^{h/2}p(Ri,z)\mathrm{d}z=hp_v \\
    N_r(R)=-\int_{-h/2}^{h/2}\frac{E_s}{h}u(R,z)\mathrm{d}z &\; , \quad 
    N_r(Ri)=\int_{-h/2}^{h/2}y\frac{E_s}{h}u(R,z)\mathrm{d}z\\
    T_r(Ri)=0 &\; , \quad w(R,0)=0\\
    M_r(R)=-\int_{-h/2}^{h/2}\frac{E_s}{h}u(R,z)z\mathrm{d}z &\; , \quad M_r(Ri)=\int_{-h/2}^{h/2}y\frac{E_s}{h}u(R,z)z\mathrm{d}z \;,
\end{align*}
where, $k_{rre}=E_s/h$ and $k_{rri}=yk_{rre}$ have been used, and $N_r, T_r, M_r$ represent the cross-sectional normal force, shear force, and bending moment, respectively, i.e.: 
\begin{eqnarray*}
    N_r(r)=\int_{-h/2}^{h/2}\sigma_{rr}\mathrm{d}z \; , \quad 
    T_r(r)=\gamma\int_{-h/2}^{h/2}\sigma_{rz}\mathrm{d}z\; , \quad   
    M_r(r)=\int_{-h/2}^{h/2}\sigma_{rr}z\ \mathrm{d}z \;.\\
\end{eqnarray*}
We note that $\gamma$ is a correction factor that is artificially introduced in Mindlin plate theory, since shear stresses are constant through the thickness as a consequence of the chosen kinematics, not reflecting the real distribution and affecting thus the total shear deformation energy. In addition, to the eight boundary conditions for the plate model, which are necessary to determine the eight integration constants of the system of PDEs governing equilibrium, two more boundary conditions are requested for the evaluation of the parameters defining the through-thickness kinematics namely, $\varepsilon_0$ and $\varepsilon_1$. The remaining conditions are those on the vertical axial stress, or
\begin{equation*}
    \int_{R_i}^{R}\int_0^{2\pi}\sigma_{zz}(r,h/2)\ r\ \mathrm{d}\theta\mathrm{d}r=-2\pi\ p_{RLT}(R^2-Ri^2) \quad , \quad  \int_{R_i}^{R}\int_0^{2\pi}\sigma_{zz}(r,-h/2)\ r\ \mathrm{d}\theta\mathrm{d}r=-2\pi\ p_{IOP}(R^2-Ri^2) \; .
\end{equation*}

\subsection{Results}
The results obtained with the proposed approach are presented below, organized into three main sections. First, a quantitative analysis is provided of the hydraulic and mechanical responses captured by the transversely isotropic LC plate model described above, under both physiological and pathological levels of intraocular pressure ($p_{IOP}$). The key plots highlight the impact of pathological conditions on the deformation and stress states of the ocular nerves, as well as on fluid content variations, which are linked to optimal tissue blood perfusion.\\
\indent  Next, since a key difference of the present approach with previous works in the literature lies in the model presented being anisotropic (transversely isotropic, TI), the question of whether the introduced anisotropy is essential to qualitatively reproduce the observed results is addressed. To this end, the proposed model is compared with a kinematically equivalent Mindlin poroelastic plate made of isotropic material. The isotropic model is obtained by leveraging the out-of-plane elastic constants of the TI formulation to those of the plane of isotropy.
As is clear from the results of the analyses, Sect. \ref{ch:effects_anisotropy}, the TI and its isotropic version show very different hydraulic and mechanical predictions, at least quantitatively, underestimating stresses and strains in regions of interest.\\
\indent Finally, a parametric study is conducted to investigate how the mechanical response of the lamina is affected by the stiffness of the opening accommodating the central retinal vessels (hereafter referred to as the central LC canal). This analysis serves both as a sensitivity study and as a means of assessing the robustness of the model, while also providing a possible route for estimating the stiffness of this structure, a parameter that has not yet been determined, either experimentally or indirectly.\\
\indent Mechanical parameters used in the model are organised in tables in the following manner: the elastic and porous parameters are summarised in Table \ref{tab:elastic}, the boundary conditions in Table \ref{tab:bc}, and the geometric quantities in Table \ref{tab:geometry}.
\begin{table}[h!]
\centering
\begin{tabular}{l l c c r}
\hline
\textbf{Description} & \textbf{Symbol} & \textbf{Units} & \textbf{Value} & \textbf{Ref.} \\
\hline
Matrix Poisson ratio & $\nu_m$ & $-$ & 0.49 & \cite{Karimi2021,Karimi2022} \\
Matrix Young's modulus  & $\mathrm{E}_m$ & $\textup{Pa}$ & 
670000 & 
\cite{Karimi2021,Karimi2022}\\
Primary porosity & $\phi$ & $-$ & 0.43 & \cite{Ling2019} \\
Poisson ratio $(r,z)$ & $\nu_z$ & $-$ & 0.49 & - \\ 
Poisson ratio $(r,\theta)$ & $\nu_t$ & $-$ & 0.351507 & - \\
Young's modulus $(r, \theta)$ & $E_t$ & $\textup{Pa}$ & 215025 & - \\
Young's modulus $z$ & $E_z$ & $\textup{Pa}$ & 381887 & - \\
Shear modulus in ($r,z$) & $\mu_z$ & $\textup{Pa}$ & 89573 &  - \\
Secondary porosity & $\varphi$ & $-$ & 0.156 & \cite{causin2014}\\
Permeability in ($r,\theta$) & $k_t$ & $\textup{m}^2$ & $1.521 \times 10^{-12}$ & \cite{causin2014} \\
Permeability in $z$ direction & $k_z$ & $\textup{m}^2$ & $1.521 \times 10^{-12}$ & \cite{causin2014} \\
Fluid viscosity & $\mu_F$ & $\textup{Pa s}$ & 0.01001 & \cite{causin2014}\\
\hline
\end{tabular}
\caption{Material parameters used in the transversely isotropic porous plate model.}
\label{tab:elastic}
\end{table}

\begin{table}[h!]
\centering
\begin{tabular}{l l c c r}
\hline
\textbf{Description} & \textbf{Symbol} & \textbf{Units} & \textbf{Value} & \textbf{Ref.} \\
\hline
Arterial pressure & $p_a$ & $\textup{mmHg}$& 30 & \cite{Prada2016} \\
Venous pressure & $p_v$ & $\textup{mmHg}$ & 20 & \cite{Prada2016} \\
Retrolaminar tissue pressure & $p_\mathrm{RLT}$ & $\textup{mmHg}$ & 7 & \cite{Prada2016} \\
Radial spring stiffness exterior & $k_{rre}$ & $\textup{N}/\textup{m}^3$ & $6.2367 \times 10^{10}$ & - \\
Radial spring stiffness interior & $k_{rri}$ & $\textup{N}/\textup{m}^3$ & $1.5592 \times 10^{10}$ & - \\
Sclera Young's modulus & $E_s$ & $\textup{Pa}$ & $1.871 \times 10^{7}$ & \cite{Grytz2011} \\
\hline
\end{tabular}
\caption{Parameters specifying the boundary conditions of the model.}
\label{tab:bc}
\end{table}

\begin{table}[h!]
\centering
\begin{tabular}{l l c c r}
\hline
\textbf{Description} & \textbf{Symbol} & \textbf{Units} & \textbf{Value} &\textbf{Ref.} \\
\hline
External radius of lamina & $R$ & $\textup{m}$ & $7.9\times 10^{-4}$ & \cite{Prada2016} \\
Central canal radius & $R_i$ & $\textup{m}$ & $1.13\times 10^{-4}$ & \cite{Prada2016} \\
Height of lamina & $h$ & $\textup{m}$ & $3\times10^{-4}$ & \cite{Prada2016} \\
\hline
\end{tabular}
\caption{Parameters specifying the geometry of the cylindrical LC of the proposed model.}
\label{tab:geometry}
\end{table}
\indent  As reported in Table \ref{tab:elastic}, the elastic parameters of the model have been selected from a strict number of studies in order to preserve the internal consistency. Indeed, with the sole exception of the primary porosity $\phi$, retrieved from \cite{Ling2019}, the dataset has been obtained from \cite{causin2014}. Since not all the engineering constants required by the present transversely isotropic poroelastic formulation were directly available in the selected literature source, the quantities that were not explicitly reported have been estimated by means of a standard numerical homogenization procedure, as reported in what follows. Although a residual uncertainty associated with biological variability inevitably remains, the present model does not aim to be subject-specific, nor does it claim that the adopted constants uniquely represent the mechanical response of the LC. Rather, its goal is to construct a mechanically coherent and analytically tractable reference model.\\
\indent  From the hemodynamic point of view, it is worth noting that in Table \ref{tab:elastic} the permeabilities $k_t$ and $k_z$ have been assumed to be equal. This is due to the fact that, although in principle the vascular organisation of the LC may suggest a more developed permeability in the in-plane directions, the adopted kinematic and static assumptions, together with no-flux conditions at the upper and lower surfaces of the LC, lead to a vanishing through-thickness fluid flux in the analytical solution. In this fashion, the choice of $k_z=k_t$ is a simplifying assumption coherent with the expected dominant fluid transport due to the vascular arrangement.
 The mechanical properties of the transversely isotropic LC, not retrievable from the literature sources available, have been derived from the experimental values available in the literature by employing
 a numerical homogenization procedure. In particular, the numerical homogenization was performed on a periodic hexagonal representative cell containing a central cylindrical void, following a standard periodic-homogenization setting \cite{day1992,parnell2009}, on Comsol Multiphysics\textsuperscript{\textregistered}.
 Starting from the solid matrix properties, assumed isotropic and obtained from \cite{Karimi2021,Karimi2022}, a hexagonal cell of side length $a$ has been built. 
To account for the high level of porosity a central cylinder of radius $b=a\sqrt{3\sqrt{3}\phi/2\pi}$, where $\phi$ is the primary porosity, has been placed. The quantity $a$, whose value does not affect the estimation of the homogenised elastic modulus, has been chosen as a small fraction of the LC radius $R$ to make $b$ physically consistent with what is observed in the scan of the LC. 
On the eight sides of the prism with hexagonal base obtained periodic boundary conditions have been imposed, while the internal cylinder has been left traction-free. Results of the analyses have been compared with self-generalised \cite{Christensen1990} estimates, and Hashin-Shtrikmann (upper) bounds, see Table \ref{tab:estimates}, to assess the physical plausibility.
\begin{table}[h!]
\centering
\begin{tabular}{l c c c c c}
\hline
\textbf{Method} & $E_z$[Pa] & $\mu_z$[Pa] & $\nu_z$[-] & $E_t$[Pa] & $\mu_t$[Pa] \\
\hline
Generalized Self-consistent,\cite{Christensen1990} & 381900 & 89618 & 0.49 & 175660 & 59718 \\
Hashin-Shtrikmann (+),\cite{hashin1965} & 381900 & 89618 & 0.49 & 230963 & 88553.3\\
Periodic Homegenization, \cite{day1992,parnell2009} & 381887 & 89573 & 0.49 & 215025 & 79550.0\\
\hline
\end{tabular}
\caption{Homogenised estimates of the elastic constants of the transversely isotropic LC obtained using different analytical (generalised self-consistent method and Hashin-Shtrikmann upper bound) and numerical (periodic homogenization on hexagonal cell) methods for an isotropic matrix ($E_m$=670000Pa, $\nu_m$=0.49 as in \cite{Karimi2021}) with cylindrical cavities and porosity $\phi=0.43$. $E_t$ and $\mu_t$ are Young's and shear moduli in the plane of isotropy of the transverse isotropic homogenised solid. $E_z$, $\mu_z$ and $\nu_z$ are the Young's modulus, shear modulus and Poisson ratio in the transverse direction, respectively.}
\label{tab:estimates}
\end{table}
\subsubsection{Effect of intraocular pressure}
\begin{figure}[!ht]
    \centering
    \includegraphics[width=\linewidth]{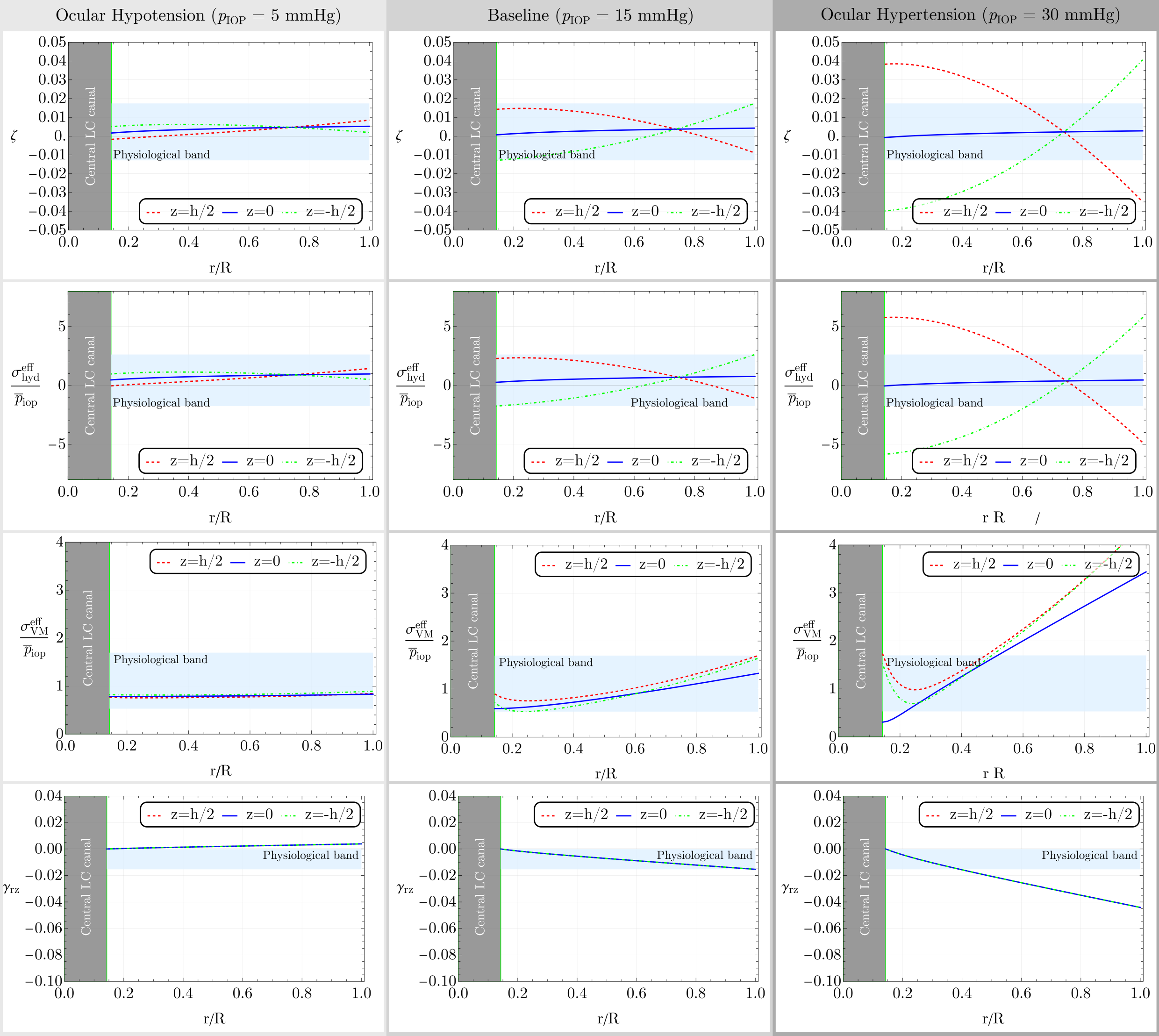}
    \caption{Variation of 
    fluid content $\zeta$ (first row), hydrostatic effective stress (which is the vascular pressure) $\sigma'_{hyd}$ (second row), Von Mises stress $\sigma'_{VM}$ (third row), and shear strain (fourth row), for three IOP conditions: ocular hypotension ($p_{IOP}=5\ \textup{mmHg}$, red, left column), baseline ($p_{IOP}=15$ mmHg, green, centre column), and ocular hypertension ($p_{IOP}=30$ mmHg, blue, right column). Stresses are normalised with respect to the baseline IOP ($\overline{p}_{IOP}=15$mmHg). Mechanical and hydraulic quantities are plotted against the normalised radial coordinate $r$ in the LC (R being the externa radius of the LC). Owing to the dominant bending behaviour of the lamina, through-thickness variations are significant and are reported using readings at the intraocular ($z=-h/2$) and retrobulbar surfaces ($z=h/2$).}
    \label{fig:variation_pressure}
\end{figure}

In this section, a quantitative analysis of the hydraulic and mechanical responses predicted by the transversely isotropic LC plate model is presented.  In what follows, the fluid component is primarily described through the variable $\zeta$, which represents the variation of the total fluid content at the homogenized scale, including both vascular and interstitial contributions. As such, $\zeta$ provides information on the local fluid availability within the tissue and may therefore be regarded as a meaningful indicator of nutrient supply and, more generally, of the local physiological state.
Strictly speaking, $\zeta$ does not represent blood perfusion alone. However, in the absence of reactive terms or inter-species exchange mechanisms, the relative proportion between vascular and interstitial fluid is assumed to remain unchanged during the deformation process. Under this assumption, variations in the total fluid content may also be interpreted as reflecting corresponding variations in the vascular contribution, and therefore as an indirect measure of the local perfusion state.
The variable $\zeta$ has been preferred here to the fluid flux $\mathbf{q}_F$ for the representation of the fluid response, since the latter follows directly and straightforwardly from the closed-form solution for the pressure and does not provide additional information on the spatial distribution of fluid-content variations. Indeed, according to Darcy's law, the fluid flux is given by
\begin{equation*}
    \mathbf{q}_F=-\frac{\mathbf{K}}{\mu_F}\nabla p(r)=-\frac{k_t}{\mu_F}\left\{p'(r),0,0\right\}.
\end{equation*}
By substituting the analytical expression of the pressure gradient, namely $p'(r)=c_1R/r$, with $c_1=(p_v-p_a)/(\log R-\log R_i)$ and where $R$ and $R_i$ denote the external and internal radii of the LC, respectively, one obtains explicitly that $\mathbf{q}_F(r)\sim 1/r$. It is worth noting that the multiplicative constant, and hence the fluid flux itself, depends on the arterial and venous pressures, $p_a$ and $p_v$, as well as on the permeability, but does not depend on $p_{IOP}$.\\
Different values of the intraocular pressure (IOP) are investigated to elucidate their effects on the mechanics of the tissue under both physiological and pathological conditions.
Fig. \ref{fig:variation_pressure} summarizes the key findings. The results are organized into three coloured columns, each showing the same hydraulic and mechanical parameters corresponding to distinct IOP levels. The physiological IOP, set at $\overline{p}_{IOP}=15$ mmHg, serves as the baseline. Two pathological conditions are also examined: ocular hypotension, represented by $p_{IOP}=5$ mmHg, and ocular hypertension, represented by $p_{IOP}=30$ mmHg.
The first row shows the variation in fluid content (
 that can be directly related to the variation of the blood content\footnote{The fluid content $\zeta$ includes both the interstitial and blood flows, but since no reactive or exchange terms are included, the relative proportions of these fluid components are assumed to remain unchanged. Thus, the variation of $\zeta$ can be also seen as the variation of the blood content.), $\zeta$. The second row presents the hydrostatic  part of the effective stress  carried by the solid skeleton of the porous medium, $\sigma_{hyd}^{eff}=\mathbf{\sigma}^{eff}:\mathbf{I}/3$, normalized to the baseline IOP.  This is the measure of the compressive stress state within the homogenized solid phase.} The third row shows a measure of the deviatoric stress (called Von Mises stress) defined as:
\begin{equation}
\sigma_{VM}^{eff}=\sqrt{\frac{3}{2}}\sqrt{\boldsymbol{\sigma}^{eff}_{\mathrm{dev}}:\boldsymbol{\sigma}^{eff}_{\mathrm{dev}}}=\sqrt{\frac{3}{2}}\ \sqrt{(\boldsymbol{\sigma}^{eff})^2:\mathbf{I} -\frac{1}{3}(\boldsymbol{\sigma}^{eff}:\mathbf{I})^2}
\end{equation}
and finally, the transverse shear strain $\gamma_{rz}=2\varepsilon_{rz}$.\\
All plots report values at the retrobulbar surface ($z=h/2$), middle plane ($z=0$), and intraocular surface ($z=-h/2$) as functions of the radial coordinate.\\
The baseline column represents the physiological condition against which pathological states are compared. For clarity, the physiological range of values has been added to the graphs of the pathological results for direct comparison.  It is worth noting that the here investigated deviatoric stress measure, i.e. Von Mises stress, must not be seen as a rigorous tissue-specific failure criterion, but rather as a scalar measure of the distortional/deviatoric part of the stress state. The choice of using this measure is motivated by the fact that, in hydrated biological tissues, the hydrostatic part of the stress is often less directly associated with distortional mechanical effects, whereas the deviatoric component is generally more informative in describing shear-related mechanical states that may be relevant for tissue deformation and damage. In this fashion, a deviatoric stress measure (namely Von Mises stress) is often used as an effective synthetic indicator in the biomechanics literature (see e.g. \cite{Marques2019,Ardatov2023,Esposito2018,Carotenuto2026}). \\
\indent A clear observation is that in the ocular hypertensive condition (right column), several stress and deformation measures markedly exceed the physiological range. In particular, elevated Von Mises stress (third row, right column) is evident in the peripheral regions of the lamina. This is mechanically linked to the high shear deformations observed in the same area (fourth row), which are likely to stress the optic nerve head (ONH) bundles, potentially leading to nerve damage and disease progression.  The high level of shear strain detected in the peripheral region is consistent with that reported in previous studies based on poroelastic models, e.g. \cite{Tatone2019}, despite the simpler and more analytically tractable framework adopted in the present work. Thus, ocular hypertension emerges as a primary candidate for initiating nerve-related damage. Notably, Von Mises stress shows little variation through the thickness, as indicated by the near-overlapping blue, green, and red curves. In contrast, 
 fluid content variation is more sensitive to the lamina's bending behavior, showing pronounced drainage at the intraocular and retrobulbar surfaces. Ocular hypertension, in particular, results in up to a $4\%$ reduction in 
 fluid content, which could impair nutrient distribution and exacerbate tissue vulnerability. While shear strain indicates pathological regimes primarily in the lamina's periphery, both hydrostatic stress and 
 fluid content variation fall outside physiological limits even in central regions.

\indent Conversely, ocular hypotension appears less mechanically demanding on the LC, as expected given the lower IOP, with most measures remaining within physiological bounds. However, from the first column of Figure \ref{fig:variation_pressure}, a shift in mechanical behaviour is still noticeable across all parameters -- even if less pronounced in the Von Mises stress. For example, hydrostatic pressure reveals a polarity inversion: at the retrobulbar lamina surface, it shifts from a negative to a positive radial gradient (and vice versa at the intraocular surface). This introduces primarily compressive deformation regimes opposite to those observed under baseline conditions (prevalently of the bending type), suggesting that ocular hypotension subjects the tissue to unfamiliar mechanical environments.
From a clinical and experimental viewpoint, anterior displacement of the lamina cribrosa under marked IOP reduction or ocular hypotony has been reported in the literature \cite{Ryu2024,Abe2015,tan2018}, making the qualitative trend predicted by the model not devoid of physiological plausibility.
\begin{figure}[!ht]
    \centering
    \includegraphics[width=\linewidth]{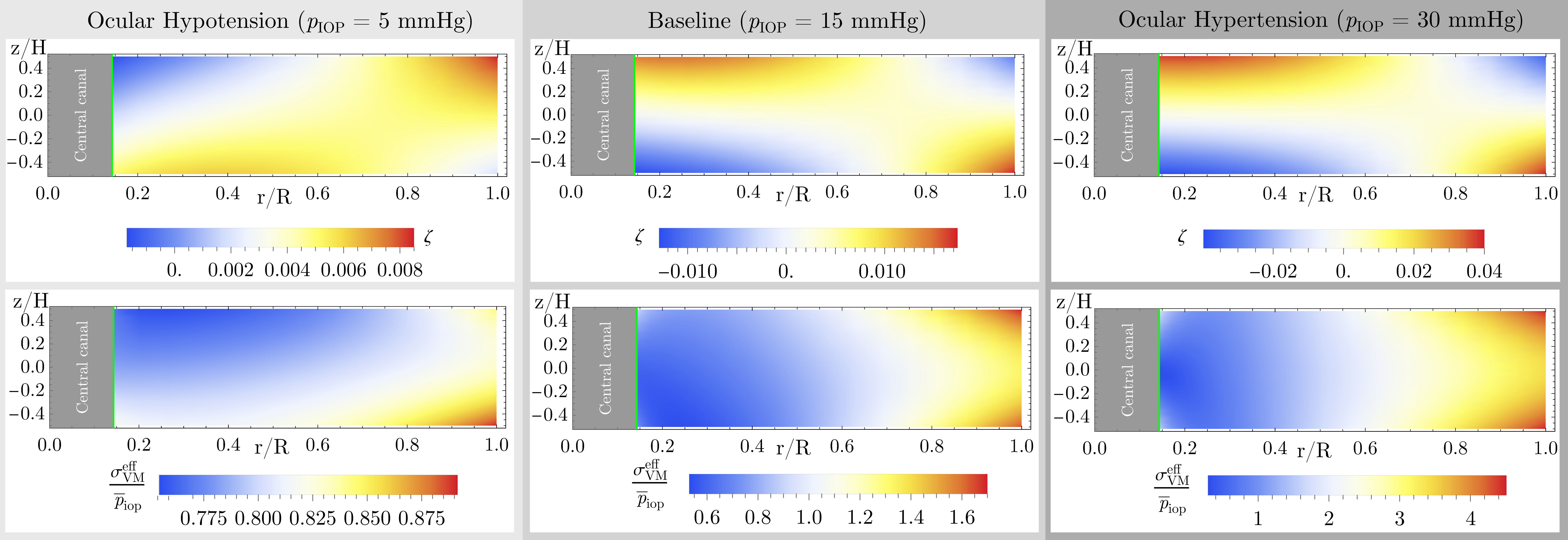}
    \caption{Color maps representing variation of fluid content ($\zeta$, first row or graphs), and a deviatoric stress measure (Von Mises stress, $\sigma_{\textup{VM}}^{\textup{eff}}$, second row of graphs), in the Mindlin poroelastic cylindrical plate model of the Lamina Cribrosa ($r$: radial direction; $z$: through-thickness direction) for different levels of intraocular pressure $p_{IOP}$ (highlighted in the columns).}
    \label{fig:placeholder}
\end{figure}
 As also reported in Figure \ref{fig:placeholder}, fluid content variation similarly reflects this regime shift, with consistently lower values compared to baseline. This indicates that in ocular hypotensive states, blood—and consequently nutrient—drainage is significantly altered at both surfaces of the lamina, irrespective of proximity to the periphery.

\subsubsection{Effects of anisotropy on the mechanics of the LC} \label{ch:effects_anisotropy}
\begin{figure}[!ht]
    \centering
    \includegraphics[width=\linewidth]{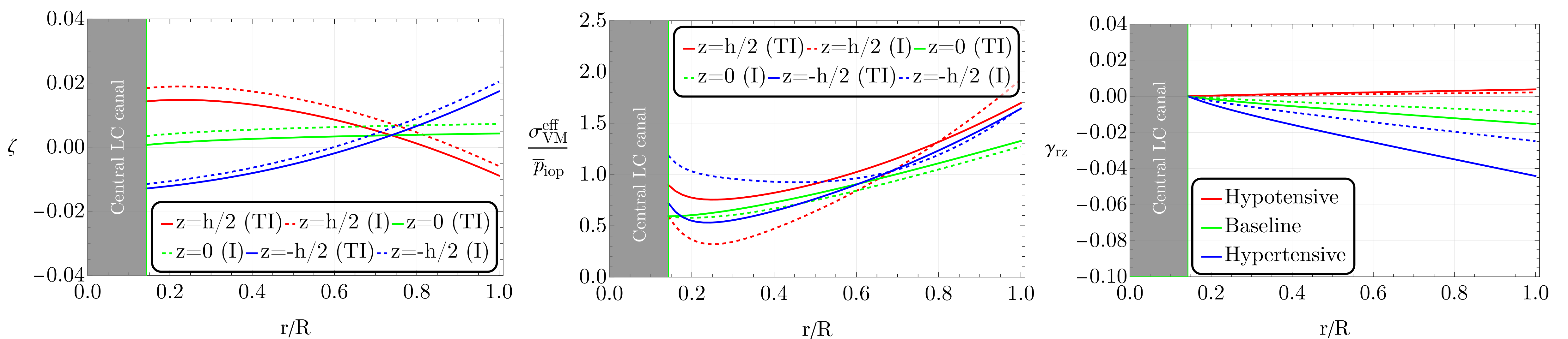}
    \caption{Comparison between transversely isotropic model (TI) and isotropic one (I), reported in terms of variation of 
     fluid} content $\zeta$ (on the left), Von Mises stress $\sigma_{VM}$ (centrally) and shear strain $\gamma_{rz}$ (on the right). Readings are reported for different values of the through-thickness coordinate.
    \label{fig:comparison_isotropy}
\end{figure}

In this section, a comparison is conducted between the transversely isotropic model presented above for the poroelastic LC and an isotropic version. For this comparison, the poroelastic independent material constants in the isotropic case (two from the elastic skeleton and two from the poroelastic interaction) are obtained from the eight independent constants of linear poroelasticity for transverse isotropy (five for the elastic skeleton and three for the porous to skeleton interaction) by leveraging the out-of-plane modulus to the one in the isotropic plane ($r,\theta$). The reason for doing so lies in the fact that in the literature \cite{causin2014,woo1972}, the elastic parameters of the lamina have been primarily determined from experimenting on the tissue along the radial direction, which corresponds to the plane of isotropy in the present TI formulation.\\
\indent The results for the case of the isotropic and the transversely isotropic cylindrical LC are reported in Figure \ref{fig:comparison_isotropy}.
Accordingly, the isotropic (I) model can follow only qualitatively the profiles of the transversal isotropic one (TI), failing at reproducing -- with error up to 60$\%$ -- the prediction of the TI model. In particular, the variation of 
 fluid content, depicted in the first graph on the left of Figure \ref{fig:comparison_isotropy}, denounces great differences between the two approaches. The consequences of using an isotropic model can then have a direct impact on the evaluation of blood inflow and outflow from the various regions of the lamina, not providing a reliable tool for the hydro-mechanical behavior of this specific tissue. This relatively large quantitative detachment between isotropy and anisotropy (TI) models is also evident from the central graph of the same figure and below in figure \ref{fig:density_iso}, where the deviatoric stress measure is reported. Here, the difference in predictions can go beyond $45\%$ for the most stressed areas, namely the intraocular and retrobulbar edges of the lamina. 
\begin{figure}[!ht]
    \centering
    \includegraphics[width=0.9\linewidth]{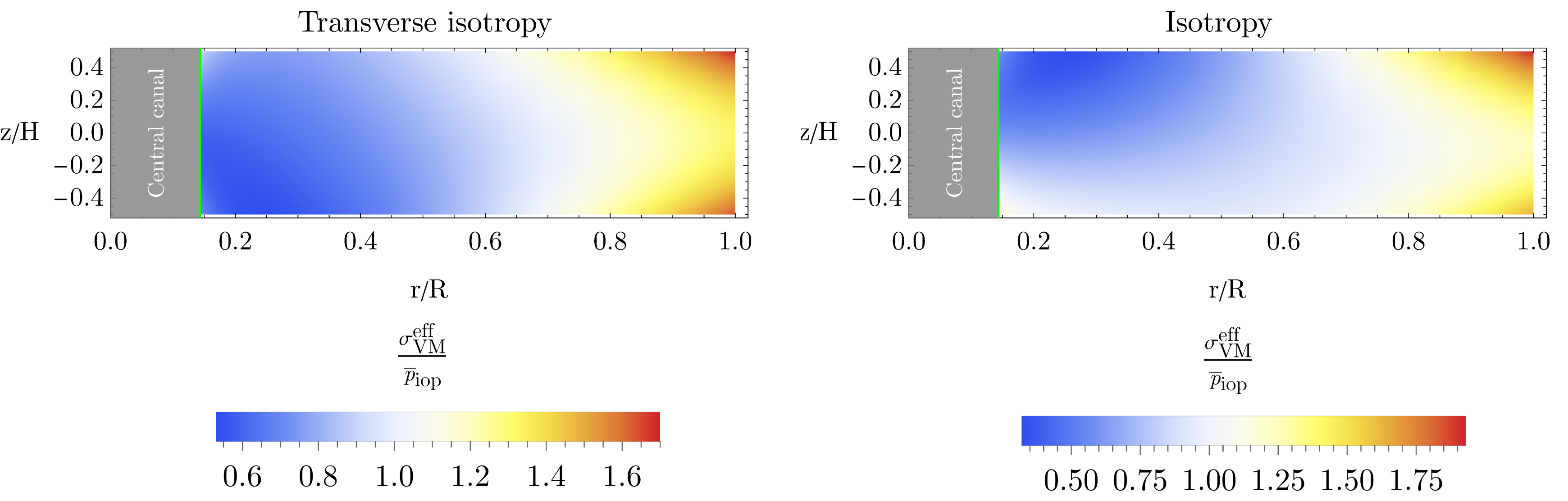}
    \caption{Density plots of a deviatoric measure, namely the normalized effective von Mises stress, through the lamina thickness, shown as a function of the dimensionless radial and thickness coordinates $r/R$ and $z/H$. The transversely isotropic model is reported on the left, while the isotropic counterpart is shown on the right. In the isotropic case, the elastic constants were taken equal to those defined in the plane of isotropy of the transversely isotropic model.}
    \label{fig:density_iso}
\end{figure}

\subsubsection{The mechanical effects of the opening for central retinal vessels}
\begin{figure}[!ht]
    \centering
    \includegraphics[width=\linewidth]{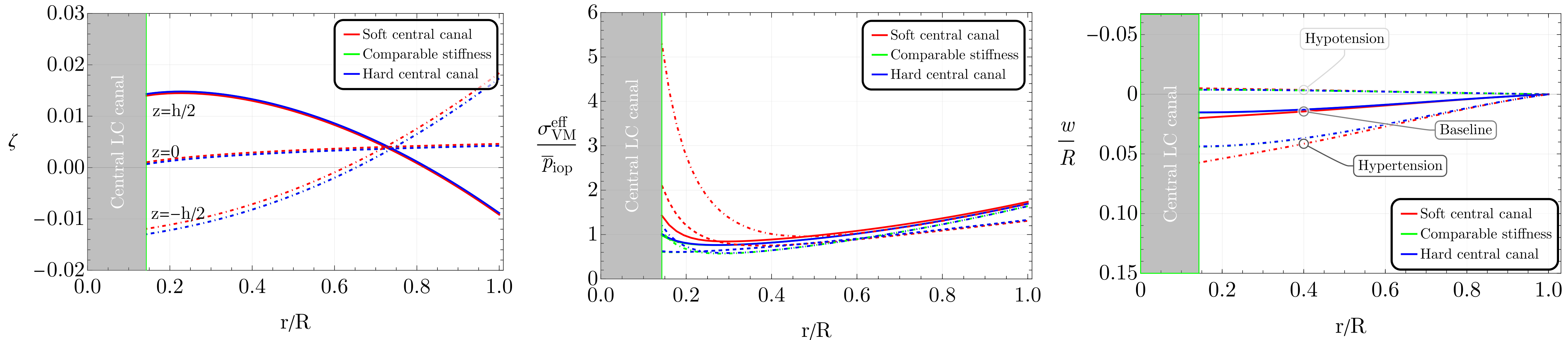}
    \caption{Analyses of the effects of the central LC canal stiffness on the poromechanics of the LC. On the left, the variation of 
     fluid} content for the basal condition $p_{IOP}=15$mmHg, given at three different heights ($z=\{-h/2,0,h/2\}$), for a soft (in red), hard (in blue) and a situation in the middle (in green) central LC canal stiffness. In the middle, the Von Mises stress is presented for the same situation of baseline inter-ocular pressure and at three different heights of the lamina for the various combination of stiffness specified above. Lastly, on the right, the effects of the central LC canal on the vertical displacements of the middle plane of the LC ($z=0$) are depicted for all three pressure-related conditions (baseline, ocular hypotension and ocular hypertension).
    \label{fig:variation_stiffness}
\end{figure}

Lastly, the influence of the central LC canal stiffness on the mechanical behavior of the LC is investigated. This analysis has been carried out since the central LC canal stiffness is a parameter that has been eluding direct evaluation for long time.\\
\indent The mechanical parameter, in the presented model, that summarizes the interaction of the LC with the canal is $k_{rr,i}$ -- the stiffness of the radial springs at $r=R_i$. To perform a parametric analysis, this quantity has been assumed to be proportional to the external radial springs stiffness $k_{rr,i}=yk_{rr,e}$ representing the interaction of the LC with the sclera in the most peripheral region of the LC.\\
\indent A comparison of the results retrieved for three different ratios of the central LC canal and the sclera ($y$) is provided in Figure \ref{fig:variation_stiffness}.
In Figure, the soft central LC canal case assumes $y\rightarrow0^+$, while the comparable stiffness case assumes $y=1$ and the hard canal case a $y\rightarrow +\infty$.
The most relevant mechanical measures to be significantly affected by the stiffness of the central LC canal are the Von Mises stress measure and the vertical displacements. The others, namely variation of 
 fluid content and 
vascular pressure are negligibly affected. In the central part of the lamina, close to the central canal, variations of up to one order of magnitude occur in the Von Mises stress while vertical displacements change up to around $35\%$, see Figure \ref{fig:variation_stiffness}. The sensitivity highlighted by this last evidence pinpoints the possibility of estimating the mechanical properties of the central LC canal from an inverse analysis using the proposed plate model and not relying on macroscopic invasive observations.

\section{Discussion and conclusions}
The results presented in this study provide new insights into the complex hydro-mechanical environment of the LC and its modulation by both IOP variations and tissue anisotropy. Under pathological IOP levels, the model highlights a scenario where mechanical stress and blood transport are tightly coupled in driving tissue health or damage. \\
Specifically, the ocular hypertensive condition emerges as the most threatening for the ONH, as evidenced by the uniformly elevated deviatoric stress measure and shear deformation across the LC thickness, particularly concentrated in the peripheral regions (see Figure \ref{fig:variation_pressure}). This stress pattern aligns with clinical observations that associate peripheral LC deformation and strain with early glaucomatous damage.
Indeed, the distribution of the strains in the LC has been investigated through a large number of clinical studies, both in-vivo \cite{Beotra2018,Czerpak2023,Czerpak2024} and ex-vivo \cite{Midgett2017}, on healthy control and glaucomatous eyes. The analyses concerned either the behavior of the LC in case of IOP lowering \cite{Czerpak2023}, hypotension case, or that related to an increase of the intraocular pressure \cite{Czerpak2024}, hypertension case. In most of these results, the shear component of the Green-Lagrangian strain tensor, as well as the maximum principal strain, and the maximum shear strain manifest higher values in the peripheral zone of the LC than closer to the central LC canal. From a mechanical point of view, these findings may be related to the connection among the lamina and the peripapillary sclera (PPS), whose anisotropic properties given by the circumferential pattern of its collagen and elastin fibers increase its possibility to resist to hoop stress \cite{Czerpak2023}. In these studies the higher values of the shear strain, maximum principal stress, and maximum shear strain have then been correlated with the thinning of the retinal nerve fiber layer (RNFL), and with a lower visual function index (VFI), showing the possible relation among the measured strains and the loss of visual field \cite{Chuangsuwanich2023}. Therefore, the present model findings not only matches the expectations in terms of deformations and strains (i.e. strains are greater at the periphery than in proximity of the central canal of the LC), but also allows the evaluation of the deviatoric stress inside the tissue which increases accordingly with the IOP toward peripheral regions of the lamina suggesting impact on RGC axons in those areas, which are thought to be responsible of the peripheral visual field.\\
The associated 4\% reduction in 
 fluid content,  a quantity proportional to blood content in the present framework, under ocular hypertensive conditions is particularly concerning, as it suggests a simultaneous compromise in tissue hypoperfusion—factors known to exacerbate neural tissue vulnerability. 
These findings reinforce the pathological role of elevated IOP not only as a mechanical aggressor but also as a disruptor of ocular tissue homeostasis. Although information on the blood content are quite difficult to obtain, some measurements have been performed with Optical Coherence Tomography Angiography (OCTA) \cite{Kim2018,Menean2025}, Laser Doppler Flowmetry \cite{Nicolela1996}, and Fluorescence \cite{Waxman2022}. 
These investigations concerned both glaucomatous, glaucomatous under hypotensive treatment, and healthy control eyes. 
The major finding of these studies is the correlation among the LC curvature index and the LC vessel density (VD), which is stronger than the association between IOP and VD \cite{Kim2018,Kim2020}. This has been studied in \cite{Kim2020}, where the ONH perfusion was investigated in treatment-na\"ive normal-tension glaucoma. For the same level of IOP, the Authors discovered that the mechanical strains in the LC were better represented by the deformation of the tissue (i.e. the curvature index of the LC), than by the level of IOP. The variations in the structure of the lamina may contribute to the deterioration of the ONH perfusion, thus confirming the importance of considering the coupling of the mechanical and hemodynamic components. The change in the compliance of the LC due to age \cite{Embleton2002} or to the pathological condition leads to higher mechanical stress and possibly to the reduction of the vessel and perfusion density. It is still not clear if this reduction is associated to the remodelling of the vascular system inside the LC or to the decrease of blood flow. In each of these cases it is supposed that this different deployment of the nutrients may lead to RGC axons death, which is also suggested by the correlation of the VD with the RNFL thinning \cite{Kim2020}. Another assumption, instead, concerns the possibility that the RGC axons loss is the leading cause of the microvascular remodelling inside the LC, and the decrease in the VD is only a consequence. 
In the present study, the results reported in the first row of Figure (\ref{fig:variation_pressure}) seem to confirm the correlation among $\zeta$ and the deformation of the tissue. 
Indeed, the retrobulbar and intraocular plane of the lamina (red and green curves in the plots of Figure \ref{fig:variation_pressure}), show a  redistribution of the blood flow in the thickness of the LC, occurring due to the deformation process.
For the cases of $p_{IOP}>p_{RLTp}$ (i.e. baseline and hypertension) blood content increases in the retrobulbar plane close to the central canal while the intraocular plane shows an increase in blodd content in the peripheral zone. In an almost symmetrical fashion, fluid (or blood, as it is assumed to keep constant concentration in fluid content) intraocular plane shows a decrease of fluid content close to the central canal while retrolaminar plane shows a decrease in the same quantity in the peripheral region.
On the other hand, when $p_{IOP}<p_{RLTp}$, i.e. ocular hypotensive regime, though often overlooked, reveals a distinct mechanical signature characterized by a reversal (when compared to baseline and hypertensive conditions) in the radial vascular pressure trend and associated deformation modes. 
While the stress magnitudes are lower, the presence of unfamiliar deformation regimes suggests that even reduced IOP levels might perturb tissue mechanics and fluid balance in ways that affect long-term structural integrity. This aspect could have implications for conditions such as ocular hypotony, where chronic low IOP has been linked to structural alterations in the ONH, despite the absence of elevated stress levels. 
These observations may suggest the influence of hemodynamics in low-tension and normal-tension glaucoma \cite{Kitazawa1986}. Indeed, this glaucomatous condition is known to be more prone to the formation of disc hemorrhages and to paracentral visual defects, that might be partially justified by the change in curvature and blood flow distribution in the LC. Overall, these results highlight that both extremes of IOP -- hypertension and hypotension -- can impose mechanical environments that deviate from the physiological norm, potentially leading to pathological remodelling.\\
\indent The comparison between anisotropic and isotropic models further underscores a key methodological implication: neglecting the LC’s anisotropy results in substantial quantitative errors—up to 60\% in blood variation and 45\% in deviatoric stress predictions. This overestimation is not merely a modelling artefact but can translate into misleading interpretations when assessing risk factors or planning interventions, such as IOP-lowering therapies. Given the LC’s known collagen fibre architecture, which imparts direction-dependent stiffness, these findings advocate for the routine incorporation of anisotropy in computational models aiming at predicting ONH biomechanics and perfusion.\\
\indent Finally, the parametric study on central LC canal stiffness uncovers a crucial, yet previously underexplored, factor influencing LC mechanics. The strong sensitivity of Von Mises stress (up to an order of magnitude variation) and vertical displacements (up to 35\%) to the canal’s stiffness parameter suggests that the mechanical characterisation of this structure is far from negligible. These results pave the way for future studies employing inverse modelling techniques to estimate canal stiffness from in vivo measurements, potentially enabling personalized diagnostic assessments of LC biomechanical health. Moreover, understanding the role of canal stiffness could inform the development of therapeutic strategies aimed at modulating local mechanical environments to mitigate stress concentrations and preserve ONH function in glaucoma and other optic neuropathies. \\
\indent  A meaningful future extension of the present work may be represented by the integration of a deformation-dependent permeability, as proposed e.g. in Tatone et al. \cite{Tatone2019}, with a precise distinction among the different fluid phases. This will require to properly consider the specific microarchitecture of the microvasculature and, accordingly, its homogenization.

\section*{Acknowledgements}
RC and MF thank the financial support from the project FIT4MEDROB, PNC0000007 (ID 62053). 
M.F. additionally thanks financial support from MUR through the project AMPHYBIA (PRIN-2022ATZCJN).\\
\indent LD and SD gratefully acknowledge the Italian Ministry of the University and Research (MUR) in the framework of the project DICAM-EXC, Departments of Excellence 2023-2027 (grant DM 230/2022), and the support from the grant PRIN-2022XLBLRX. LD and SD also gratefully acknowledge the partial support from the following funding sources: 2023-2025 PNRR\_CN\_ICSC\_Spoke 7\_CUP E63C22000970007 grant from the Italian Government,  ERC-CG S-FOAM Horizon Europe EU\_HE\_GA 101086644, ERC-ADG-2021-101052956-BEYOND,  SUBBIMATT Horizon Europe EU\_HE\_GA 101129911.\\
\indent GG thanks financial support from the National Eye Institute through the project SCH: SEE through GLAUCOMA: Smart Eye Emulator (SEE) to study glaucoma risk factors, NIH grant R01EY034718.\\
\indent AH is supported by NIH grant R01EY034718, NYEE Foundation grants, The Glaucoma Foundation, and in part by a Challenge Grant award from Research to Prevent Blindness, NY.  AH is supported by the Barry Family Center for Ophthalmic Artificial Intelligence \& Human Health.\\

This work was also conducted under the auspices of GNFM-INDAM, founded by the Italian Ministry of Universities and Research.

\section*{Authors' contributions}
MF, LD, GG and AH conceptualized, coordinated, and edited the work. All the authors contributed to the writing of the manuscript.
SB and AV reviewed the manuscript and took part in the writing of Sect. 1 and 4.
LD and GG reviewed the manuscript and contributed to the writing.
SD contributed to the writing of the manuscript.
MF, RC and SD developed the mechanical model. RC and SD performed the analyses and created the figures.
All the authors have read and approved the final manuscript.

\section*{Conflict of Interests}
Professor Alon Harris would like to disclose that he received remuneration from AdOM, Qlaris, and Cipla for serving as a consultant, and he serves on the board of AdOM, Qlaris and SlitLed. Professor Alon Harris holds an ownership interest in AdOM, Oxymap, Qlaris, SlitLed, and  AEYE Health. If you have questions regarding paid relationships that your physician/researcher may have with industry, you are encouraged to talk with your physician/researcher, or check for industry relationships posted on individual faculty pages on our website athttp://icahn.mssm.edu/.

\bibliographystyle{elsarticle-num}
\bibliography{bibMorph}

@article{Alaimo2019FractionalPoroelasticity,
title = {A fractional order theory of poroelasticity},
journal = {Mechanics Research Communications},
volume = {100},
pages = {103395},
year = {2019},
issn = {0093-6413},
doi = {https://doi.org/10.1016/j.mechrescom.2019.103395},
url = {https://www.sciencedirect.com/science/article/pii/S009364131930103X},
author = {G. Alaimo and V. Piccolo and A. Cutolo and L. Deseri and M. Fraldi and M. Zingales}
}

@Article{sladek2015,
  author    = {J. Sladek and V. Sladek and M. Gfrerer and M. Schanz},
  journal   = {Acta Mech},
  title     = {Mindlin theory for the bending of porous plates},
  year      = {2015},
  volume    = {226},
  pages     = {1909--1928},
  publisher = {Springer-Verlag Wien},
}

@Article{woo1972,
  author    = {P. I-Y. Woo and A.S. Kobayashi and W.A. Schlegel and C. Lawrence},
  journal   = {Exp. Eye Res.},
  title     = {Nonlinear Material Properties of Intact Cornea and Sclera},
  year      = {1972},
  volume    = {14},
  pages     = {29--39}
}

@book{COUSSY2004,
	author = {Olivier Coussy},
	publisher = {John Wiley \& Sons, Ltd},
	title = {Poromechancis},
    city = {Chichester, West Sussex},
	year = {2004}
}

@book{COWIN2007,
    author = {Stephen C. Cowin and Stephen B. Doty},
    title = {Tissue Mechanics},
    publisher = {Springer New York, NY},
    year = {2007}
}

@book{Wang2000,
    author = {F.H. Wang},
    title = {Theory of Linear Poroelasticity with Applications to Geomechanics and Hydrogeology},
    publisher = {Princeton University Press},
    year = {2001},
    city = {Princeton}
}

@article{BIOT1941,
title = {General Theory of Three‐Dimensional Consolidation},
journal = {J. Appl. Phys.},
volume = {12},
issue = {2},
pages = {155--164},
year = {1941},
author = {Maurice A. Biot}
}

@article{FRALDI2018,
    author = {M. Fraldi and A.R. Carotenuto},
    title = {Cells competition in tumor growth poroelasticity},
    journal = {Journal of the Mechanics and Physics of Solids},
    year = {2018},
    volume = {112},
    pages = {345--367}
}

@article{CAROTENUTO2021,
    author = {Angelo Rosario Carotenuto and Arsenio Cutolo and Stefania Palumbo and Massimiliano Fraldi},
    title = {Lyapunov stability of competitive cells dynamics in tumor mechanobiology},
    journal = {Acta Mechanica Sinica},
    year = {2021},
    volume = {37},
    issue = {2},
    pages = {244–-263}
}

@article{causin2014,
    author = {Paola Causin and Giovanna Guidoboni and Alon Harris and Daniele Prada and Riccardo Sacco and Samuele Terragni},
    title = {A poroelastic model for the perfusion of the lamina cribrosa in the optic nerve head},
    journal = {Mathematical Biosciences},
    year = {2014},
    volume = {257},
    pages = {33--41}
}

@phdthesis{Prada2016,
    author = {Daniele Prada},
    title = {A Hybridizable Discontinuous Galerkin Method for Nonlinear Porous Media Viscoelasticity with Applications in Ophtalmology},
    school = {Purdue University, Indianapolis, Indiana, USA},
    year = {2016}
}

@article{Grytz2011,
    title = {The collagen fibril architecture in the lamina cribrosa and peripapillary sclera predicted by a computational remodeling approach},
    journal = {Biomechanics and Modeling in Mechanobiology},
    volume = {10},
    pages = {371-382},
    year = {2011},
    doi = {https://doi.org/10.1007/s10237-010-0240-8},
    author = {Rafael Grytz and G\"unther Meschke and Jost B. Jonas}
}

@article{Ling2019,
    author = {Ling, Yik Tung Tracy and Shi, Ran and Midgett, Dan E. and Jefferys, Joan L. and Quigley, Harry A. and Nguyen, Thao D.},
    title = {Characterizing the Collagen Network Structure and Pressure-Induced Strains of the Human Lamina Cribrosa},
    journal = {Investigative Ophthalmology \& Visual Science},
    volume = {60},
    number = {7},
    pages = {2406-2422},
    year = {2019},
    month = {06},
    issn = {1552-5783},
    doi = {10.1167/iovs.18-25863},
    url = {https://doi.org/10.1167/iovs.18-25863},
    eprint = {https://arvojournals.org/arvo/content\_public/journal/iovs/938044/i1552-5783-60-7-2406.pdf},
}

@article{Beotra2018,
    author = {Beotra, Meghna R. and Wang, Xiaofei and Tun, Tin A. and Zhang, Liang and Baskaran, Mani and Aung, Tin and Strouthidis, Nicholas G. and Girard, Michaël J. A.},
    title = {In Vivo Three-Dimensional Lamina Cribrosa Strains in Healthy, Ocular Hypertensive, and Glaucoma Eyes Following Acute Intraocular Pressure Elevation},
    journal = {Investigative Ophthalmology \& Visual Science},
    volume = {59},
    number = {1},
    pages = {260-272},
    year = {2018},
    month = {01},
    issn = {1552-5783},
    doi = {10.1167/iovs.17-21982},
    url = {https://doi.org/10.1167/iovs.17-21982},
    eprint = {https://arvojournals.org/arvo/content\_public/journal/iovs/936670/i1552-5783-59-1-260.pdf},
}

@article{Czerpak2023,
    title = {The Strain Response to Intraocular Pressure Decrease in the Lamina Cribrosa of Patients with Glaucoma},
    journal = {Ophthalmology Glaucoma},
    volume = {6},
    number = {1},
    pages = {11-22},
    year = {2023},
    issn = {2589-4196},
    doi = {https://doi.org/10.1016/j.ogla.2022.07.005},
    url = {https://www.sciencedirect.com/science/article/pii/S258941962200120X},
    author = {Cameron A. Czerpak and Michael Saheb Kashaf and Brandon K. Zimmerman and Harry A. Quigley and Thao D. Nguyen},
}

@article{Midgett2017,
    title = {The pressure-induced deformation response of the human lamina cribrosa: Analysis of regional variations},
    journal = {Acta Biomaterialia},
    volume = {53},
    pages = {123-139},
    year = {2017},
    issn = {1742-7061},
    doi = {https://doi.org/10.1016/j.actbio.2016.12.054},
    url = {https://www.sciencedirect.com/science/article/pii/S1742706116307358},
    author = {Dan E. Midgett and Mary E. Pease and Joan L. Jefferys and Mohak Patel and Christian Franck and Harry A. Quigley and Thao D. Nguyen},
}

@article{Czerpak2024,
    author = {Czerpak, Cameron A. and Kashaf, Michael Saheb and Zimmerman, Brandon K. and Mirville, Rebecca and Gasquet, Nicolas C. and Quigley, Harry A. and Nguyen, Thao D.},
    title = {The Strain Response to Intraocular Pressure Increase in the Lamina Cribrosa of Control Subjects and Glaucoma Patients},
    journal = {Translational Vision Science \& Technology},
    volume = {13},
    number = {12},
    pages = {7-7},
    year = {2024},
    month = {12},
    issn = {2164-2591},
    doi = {10.1167/tvst.13.12.7},
    url = {https://doi.org/10.1167/tvst.13.12.7},
    eprint = {https://arvojournals.org/arvo/content\_public/journal/tvst/938698/i2164-2591-13-12-7\_1733304300.30537.pdf},
}

@article{Kim2020,
    author = {Kim, Ji-Ah and Kim, Tae-Woo and Lee, Eun Ji and Girard, Michael J A and Mari, Jean Martial},
    title = {Relationship between lamina cribrosa curvature and the microvasculature in treatment-na{\"\i}ve eyes},
    volume = {104},
    number = {3},
    pages = {398--403},
    year = {2020},
    doi = {10.1136/bjophthalmol-2019-313996},
    publisher = {BMJ Publishing Group Ltd},
    issn = {0007-1161},
    URL = {https://bjo.bmj.com/content/104/3/398},
    eprint = {https://bjo.bmj.com/content/104/3/398.full.pdf},
    journal = {British Journal of Ophthalmology}
}

@article{Kim2018,
    author = {Kim, Ji-Ah and Kim, Tae-Woo and Lee, Eun Ji and Girard, Michaël J. A. and Mari, Jean Martial},
    title = {Microvascular Changes in Peripapillary and Optic Nerve Head Tissues After Trabeculectomy in Primary Open-Angle Glaucoma},
    journal = {Investigative Ophthalmology \& Visual Science},
    volume = {59},
    number = {11},
    pages = {4614-4621},
    year = {2018},
    month = {09},
    issn = {1552-5783},
    doi = {10.1167/iovs.18-25038},
    url = {https://doi.org/10.1167/iovs.18-25038},
    eprint = {https://arvojournals.org/arvo/content\_public/journal/iovs/937492/i1552-5783-59-11-4614.pdf},
}

@article{Embleton2002,
    author = {Embleton, S.J. and Hosking, S.L. and Roff Hilton, E.J. and Cunliffe, I.A.},
    title ={Effect of senescence on ocular blood flow in the retina, neuroretinal rim and lamina cribrosa, using scanning laser Doppler flowmetry},
    journal = {Eye},
    volume = {16},
    pages = {156-162},
    year = {2002},
    URL = {https://doi.org/10.1038/sj.eye.6700100}
}

@article{Nicolela1996,
    title = {Scanning Laser Doppler Flowmeter Study of Retinal and Optic Disk Blood Flow in Glaucomatous Patients},
    journal = {American Journal of Ophthalmology},
    volume = {122},
    number = {6},
    pages = {775-783},
    year = {1996},
    issn = {0002-9394},
    doi = {https://doi.org/10.1016/S0002-9394(14)70373-3},
    url = {https://www.sciencedirect.com/science/article/pii/S0002939414703733},
    author = {Marcelo T. Nicolela and Peter Hnik and Stephen M. Drance}
}

@article{Menean2025,
    author = {Matteo Menean and Lorenzo Bianco and Lida Perna and Gaia L'Abbate and Rosangela Lattenzio and Francesco Bandello and Luisa Pierro},
    title ={Lamina cribrosa perfusion density is reduced in eyes with central retinal vein occlusion},
    journal = { Graefe's Archive for Clinical and Experimental Ophthalmology },
    year = {2025},
    URL = {https://doi.org/10.1007/s00417-025-06853-2}
}

@article{Waxman2022,
    title = {Lamina cribrosa vessel and collagen beam networks are distinct},
    journal = {Experimental Eye Research},
    volume = {215},
    pages = {108916},
    year = {2022},
    issn = {0014-4835},
    doi = {https://doi.org/10.1016/j.exer.2021.108916},
    url = {https://www.sciencedirect.com/science/article/pii/S0014483521004826},
    author = {Susannah Waxman and Bryn L. Brazile and Bin Yang and Po-Yi Lee and Yi Hua and Alexandra L. Gogola and Po Lam and Andrew P. Voorhees and Joseph F. Rizzo and Tatjana C. Jakobs and Ian A. Sigal}
}

@article{Bellezza2003,
    author = {Bellezza, Anthony J. and Rintalan, Christopher J. and Thompson, Hilary W. and Downs, J. Crawford and Hart, Richard T. and Burgoyne, Claude F.},
    title = {Deformation of the Lamina Cribrosa and Anterior Scleral Canal Wall in Early Experimental Glaucoma},
    journal = {Investigative Ophthalmology \& Visual Science},
    volume = {44},
    number = {2},
    pages = {623-637},
    year = {2003},
    month = {02},
    issn = {1552-5783},
    doi = {10.1167/iovs.01-1282},
    url = {https://doi.org/10.1167/iovs.01-1282},
    eprint = {https://arvojournals.org/arvo/content\_public/journal/iovs/932919/7g0203000623.pdf},
}

@article{Quigley1995,
    author = {Quigley, Harry A.},
    title = {Ganglion cell death in glaucoma: pathology recapitulates ontogeny},
    journal = {Australian and New Zealand Journal of Ophthalmology},
    volume = {23},
    number = {2},
    pages = {85-91},
    doi = {https://doi.org/10.1111/j.1442-9071.1995.tb00135.x},
    url = {https://onlinelibrary.wiley.com/doi/abs/10.1111/j.1442-9071.1995.tb00135.x},
    eprint = {https://onlinelibrary.wiley.com/doi/pdf/10.1111/j.1442-9071.1995.tb00135.x},
    year = {1995}
}

@article{Anderson1974,
    author = {Anderson, Douglas R. and Hendrickson, Anita},
    title = {Effect of Intraocular Pressure on Rapid Axoplasmic Transport in Monkey Optic Nerve},
    journal = {Investigative Ophthalmology \& Visual Science},
    volume = {13},
    number = {10},
    pages = {771-783},
    year = {1974},
    month = {10},
    issn = {1552-5783},
    eprint = {https://arvojournals.org/arvo/content\_public/journal/iovs/932880/771.pdf},
}

@article{Weber1998,
    author = {Weber, A J and Kaufman, P L and Hubbard, W C},
    title = {Morphology of single ganglion cells in the glaucomatous primate retina.},
    journal = {Investigative Ophthalmology \& Visual Science},
    volume = {39},
    number = {12},
    pages = {2304-2320},
    year = {1998},
    month = {11},
    issn = {1552-5783},
    eprint = {https://arvojournals.org/arvo/content\_public/journal/iovs/933203/2304.pdf},
}

@article{Burgoyne2011,
    title = {A biomechanical paradigm for axonal insult within the optic nerve head in aging and glaucoma},
    journal = {Experimental Eye Research},
    volume = {93},
    number = {2},
    pages = {120-132},
    year = {2011},
    note = {What Damages Ganglion Cells in Glaucoma? A Tribute to M. Rosario Hernandez},
    issn = {0014-4835},
    doi = {https://doi.org/10.1016/j.exer.2010.09.005},
    url = {https://www.sciencedirect.com/science/article/pii/S0014483510003003},
    author = {Claude F. Burgoyne},
}

@article{Voorhees2017,
    title = {Effects of collagen microstructure and material properties on the deformation of the neural tissues of the lamina cribrosa},
    journal = {Acta Biomaterialia},
    volume = {58},
    pages = {278-290},
    year = {2017},
    issn = {1742-7061},
    doi = {https://doi.org/10.1016/j.actbio.2017.05.042},
    url = {https://www.sciencedirect.com/science/article/pii/S1742706117303343},
    author = {A.P. Voorhees and N.-J. Jan and I.A. Sigal}
}

@article{Sander2006,
    author = {Sander, E. A. and Downs, J. C. and Hart, R. T. and Burgoyne, C. F. and Nauman, E. A.},
    title = {A Cellular Solid Model of the Lamina Cribrosa: Mechanical Dependence on Morphology},
    journal = {Journal of Biomechanical Engineering},
    volume = {128},
    number = {6},
    pages = {879-889},
    year = {2006},
    month = {06},
    issn = {0148-0731},
    doi = {10.1115/1.2354199},
    url = {https://doi.org/10.1115/1.2354199},
    eprint = {https://asmedigitalcollection.asme.org/biomechanical/article-pdf/128/6/879/5519087/879\_1.pdf},
}

@article{Sigal2004,
    author = {Sigal, Ian A. and Flanagan, John G. and Tertinegg, Inka and Ethier, C. Ross},
    title = {Finite Element Modeling of Optic Nerve Head Biomechanics},
    journal = {Investigative Ophthalmology \& Visual Science},
    volume = {45},
    number = {12},
    pages = {4378-4387},
    year = {2004},
    month = {12},
    issn = {1552-5783},
    doi = {10.1167/iovs.04-0133},
    url = {https://doi.org/10.1167/iovs.04-0133},
    eprint = {https://arvojournals.org/arvo/content\_public/journal/iovs/933228/z7g01204004378.pdf},
}

@article{Ayyalasomayajula2016,
	author = {Avinash Ayyalasomayajula and Robert I. Park and Bruce R. Simon and Jonathan P. Vande Geest},
	title = {A porohyperelastic finite element model of the eye: the influence of stiffness and permeability on intraocular pressure and optic nerve head biomechanics},
	journal = {Computer Methods in Biomechanics and Biomedical Engineering},
	volume = {19},
	number = {6},
	pages = {591--602},
	year = {2016},
	publisher = {Taylor \& Francis},
	doi = {10.1080/10255842.2015.1052417},
	note ={PMID: 26195024},
	URL = {https://doi.org/10.1080/10255842.2015.1052417},
	eprint = {https://doi.org/10.1080/10255842.2015.1052417}
}

@article{Edwards2001,
	author = {Michael E. Edwards and Theresa A. Good},
	title = {Use of a mathematical model to estimate stress and strain during elevated pressure induced lamina cribrosa deformation},
	journal = {Current Eye Research},
	volume = {23},
	number = {3},
	pages = {215--225},
	year = {2001},
	publisher = {Taylor \& Francis},
	doi = {10.1076/ceyr.23.3.215.5460},
	note ={PMID: 11803484},
	URL = {https://doi.org/10.1076/ceyr.23.3.215.5460},
	eprint = {https://doi.org/10.1076/ceyr.23.3.215.5460}
}

@article{Dongqi1999,
    title = {A biomathematical model for pressure-dependent lamina cribrosa behavior},
    journal = {Journal of Biomechanics},
    volume = {32},
    number = {6},
    pages = {579-584},
    year = {1999},
    issn = {0021-9290},
    doi = {https://doi.org/10.1016/S0021-9290(99)00025-1},
    url = {https://www.sciencedirect.com/science/article/pii/S0021929099000251},
    author = {He Dongqi and Ren Zeqin}
}

@article{Mindlin1951,
    author = {Mindlin, R. D.},
    title = {Influence of Rotatory Inertia and Shear on Flexural Motions of Isotropic, Elastic Plates},
    journal = {Journal of Applied Mechanics},
    volume = {18},
    number = {1},
    pages = {31-38},
    year = {1951},
    month = {04},
    issn = {0021-8936},
    doi = {10.1115/1.4010217},
    url = {https://doi.org/10.1115/1.4010217},
    eprint = {https://asmedigitalcollection.asme.org/appliedmechanics/article-pdf/18/1/31/6747060/31\_1.pdf},
}

@article{Reissner1945,
    author = {Reissner, Eric},
    title = {The Effect of Transverse Shear Deformation on the Bending of Elastic Plates},
    journal = {Journal of Applied Mechanics},
    volume = {12},
    number = {2},
    pages = {A69-A77},
    year = {1945},
    month = {03},
    issn = {0021-8936},
    doi = {10.1115/1.4009435},
    url = {https://doi.org/10.1115/1.4009435},
    eprint = {https://asmedigitalcollection.asme.org/appliedmechanics/article-pdf/12/2/A69/6744958/a69\_1.pdf},
}

@article{Roberts2009,
    author = {Roberts, Michael D. and Grau, Vicente and Grimm, Jonathan and Reynaud, Juan and Bellezza, Anthony J. and Burgoyne, Claude F. and Downs, J. Crawford},
    title = {Remodeling of the Connective Tissue Microarchitecture of the Lamina Cribrosa in Early Experimental Glaucoma},
    journal = {Investigative Ophthalmology \& Visual Science},
    volume = {50},
    number = {2},
    pages = {681-690},
    year = {2009},
    month = {02},
    issn = {1552-5783},
    doi = {10.1167/iovs.08-1792},
    url = {https://doi.org/10.1167/iovs.08-1792},
    eprint = {https://arvojournals.org/arvo/content\_public/journal/iovs/933448/z7g00209000681.pdf},
}

@article{Geijer1979,
  title={Effects of raised intraocular pressure on retinal, prelaminar, laminar, and retrolaminar optic nerve blood flow in monkeys.},
  author={Geijer, Christoffer and Bill, Anders},
  journal={Investigative ophthalmology \& visual science},
  volume={18},
  number={10},
  pages={1030--1042},
  year={1979},
  publisher={The Association for Research in Vision and Ophthalmology}
}

@article{Fechtner1994,
    title = {Mechanisms of optic nerve damage in primary open angle glaucoma},
    journal = {Survey of Ophthalmology},
    volume = {39},
    number = {1},
    pages = {23-42},
    year = {1994},
    issn = {0039-6257},
    doi = {https://doi.org/10.1016/S0039-6257(05)80042-6},
    url = {https://www.sciencedirect.com/science/article/pii/S0039625705800426},
    author = {Robert D. Fechtner and Robert N. Weinreb}
}

@article{Kitazawa1986,
    title = {Optic Disc Hemorrhage in Low-tension Glaucoma},
    journal = {Ophthalmology},
    volume = {93},
    number = {6},
    pages = {853-857},
    year = {1986},
    issn = {0161-6420},
    doi = {https://doi.org/10.1016/S0161-6420(86)33658-3},
    url = {https://www.sciencedirect.com/science/article/pii/S0161642086336583},
    author = {Yoshiaki Kitazawa and Shiroaki Shirato and Tetsuya Yamamoto}
}

@article{Hannay2024,
    title = {A Noninvasive Clinical Method to Measure in Vivo Mechanical Strains of the Lamina Cribrosa by OCT},
    journal = {Ophthalmology Science},
    volume = {4},
    number = {4},
    pages = {100473},
    year = {2024},
    issn = {2666-9145},
    doi = {https://doi.org/10.1016/j.xops.2024.100473},
    url = {https://www.sciencedirect.com/science/article/pii/S2666914524000095},
    author = {Vanessa Hannay and Cameron Czerpak and Harry A. Quigley and Thao D. Nguyen}
}

@article{Midgett2019,
    title = {In vivo characterization of the deformation of the human optic nerve head using optical coherence tomography and digital volume correlation},
    journal = {Acta Biomaterialia},
    volume = {96},
    pages = {385-399},
    year = {2019},
    issn = {1742-7061},
    doi = {https://doi.org/10.1016/j.actbio.2019.06.050},
    url = {https://www.sciencedirect.com/science/article/pii/S1742706119304726},
    author = {Dan E. Midgett and Harry A. Quigley and Thao D. Nguyen}
}

@article{Foster2005,
  title={The impact of Vision 2020 on global blindness},
  author={Foster, Allen and Resnikoff, Serge},
  journal={Eye},
  volume={19},
  number={10},
  pages={1133--1135},
  year={2005},
  publisher={Nature Publishing Group}
}

@article{Flaxman2017,
  title={Global causes of blindness and distance vision impairment 1990--2020: a systematic review and meta-analysis},
  author={Flaxman, Seth R and Bourne, Rupert RA and Resnikoff, Serge and Ackland, Peter and Braithwaite, Tasanee and Cicinelli, Maria V and Das, Aditi and Jonas, Jost B and Keeffe, Jill and Kempen, John H and others},
  journal={The Lancet Global Health},
  volume={5},
  number={12},
  pages={e1221--e1234},
  year={2017},
  publisher={Elsevier}
}

@article{Chuangsuwanich2023,
    title = {Differing Associations between Optic Nerve Head Strains and Visual Field Loss in Patients with Normal- and High-Tension Glaucoma},
    journal = {Ophthalmology},
    volume = {130},
    number = {1},
    pages = {99-110},
    year = {2023},
    issn = {0161-6420},
    doi = {https://doi.org/10.1016/j.ophtha.2022.08.007},
    url = {https://www.sciencedirect.com/science/article/pii/S0161642022006224},
    author = {Thanadet Chuangsuwanich and Tin A. Tun and Fabian A. Braeu and Xiaofei Wang and Zhi Yun Chin and Satish Kumar Panda and Martin Buist and Nicholas Strouthidis and Shamira Perera and Monisha Nongpiur and Tin Aung and Michaël J.A. Girard}
}

@article{Bernard2024,
    author = {Bernard, Chiara and Carotenuto, Angelo Rosario and Pugno, Nicola Maria and Fraldi, Massimiliano and Deseri, Luca},
    title = {Modelling lipid rafts formation through chemo-mechanical interplay triggered by receptor–ligand binding},
    journal = {Biomech Model Mechanobiol},
    volume ={23},
    year = {2024},
    pages = {485-505}
}

@article{Marques2019,
  author = {Marques, M. and Belinha, J. and Dinis, L. M. J. S. and Natal Jorge, R. M.},
  title = {A new numerical approach to mechanically analyse biological structures},
  journal = {Computer Methods in Biomechanics and Biomedical Engineering},
  volume = {22},
  number = {1},
  pages = {100--111},
  year = {2019}
}

@article{Carotenuto2026,
  author = {Carotenuto, A. R. and Cutolo, A. and Bernard, C. and others},
  title = {Adaptation-driven optimization strategies in reinforced vascular autografts},
  journal = {Meccanica},
  volume = {61},
  pages = {40},
  year = {2026},
  doi = {10.1007/s11012-026-02100-y}
}

@article{Ardatov2023,
  author = {Ardatov, O. and Aleksiuk, V. and Maknickas, A. and Stonkus, R. and Uzieliene, I. and Vaiciuleviciute, R. and Pachaleva, J. and Kvederas, G. and Bernotiene, E.},
  title = {Modeling the Impact of Meniscal Tears on von Mises Stress of Knee Cartilage Tissue},
  journal = {Bioengineering},
  volume = {10},
  number = {3},
  pages = {314},
  year = {2023},
  doi = {10.3390/bioengineering10030314}
}

@article{Esposito2018,
    author = {L. Esposito and P. Bifulco and P. Gargiulo and M.K. Gíslason and M. Cesarelli and L. Iuppariello and H. Jónsson and A. Cutolo and M. Fraldi},
    title = {Towards a patient-specific estimation of intra-operative femoral fracture risk},
    journal = {Computer Methods in Biomechanics and Biomedical Engineering},
    volume = {21},
    number = {12},
    pages = {663--672},
    year = {2018},
    publisher = {Taylor \& Francis},
    doi = {10.1080/10255842.2018.1508570},
    note ={PMID: 30370789},
    URL = {https://doi.org/10.1080/10255842.2018.1508570},
    eprint = {https://doi.org/10.1080/10255842.2018.1508570}
}

@article{Christensen1990,
    title = {A critical evaluation for a class of micro-mechanics models},
    journal = {Journal of the Mechanics and Physics of Solids},
    volume = {38},
    number = {3},
    pages = {379-404},
    year = {1990},
    issn = {0022-5096},
    doi = {https://doi.org/10.1016/0022-5096(90)90005-O},
    url = {https://www.sciencedirect.com/science/article/pii/002250969090005O},
    author = {Richard M. Christensen}
}

@article{Karimi2021,
    title = {Modeling the biomechanics of the lamina cribrosa microstructure in the human eye},
    journal = {Acta Biomaterialia},
    volume = {134},
    pages = {357-378},
    year = {2021},
    issn = {1742-7061},
    doi = {https://doi.org/10.1016/j.actbio.2021.07.010},
    url = {https://www.sciencedirect.com/science/article/pii/S1742706121004414},
    author = {Alireza Karimi and Seyed Mohammadali Rahmati and Rafael G. Grytz and Christopher A. Girkin and J. Crawford Downs}
}

@article{Karimi2022,
    author = {Karimi, Alireza and Razaghi, Reza and Rahmati, Seyed Mohammadali and Girkin, Christopher A. and Downs, J. Crawford},
    title = {Relative Contributions of Intraocular and Cerebrospinal Fluid Pressures to the Biomechanics of the Lamina Cribrosa and Laminar Neural Tissues},
    journal = {Investigative Ophthalmology \& Visual Science},
    volume = {63},
    number = {11},
    pages = {14-14},
    year = {2022},
    month = {10},
    issn = {1552-5783},
    doi = {10.1167/iovs.63.11.14},
    url = {https://doi.org/10.1167/iovs.63.11.14}
}

@article{Li2020,
  title={Biomechanical research into lamina cribrosa in glaucoma},
  author={Li, Long and Song, Fan},
  journal={National science review},
  volume={7},
  number={8},
  pages={1277--1279},
  year={2020},
  publisher={Oxford University Press}
}

@article{Tatone2019,
    title = {From species diffusion to poroelasticity and the modeling of lamina cribrosa},
    journal = {Journal of the Mechanics and Physics of Solids},
    volume = {124},
    pages = {849-870},
    year = {2019},
    issn = {0022-5096},
    doi = {https://doi.org/10.1016/j.jmps.2018.11.017},
    url = {https://www.sciencedirect.com/science/article/pii/S0022509618302473},
    author = {A. Tatone and F. Recrosi and R. Repetto and G. Guidoboni}
}

@article{Cavuoto2026,
    title = {Poroelasticity in the presence of active fluids},
    journal = {International Journal of Engineering Science},
    volume = {221},
    pages = {104457},
    year = {2026},
    issn = {0020-7225},
    doi = {https://doi.org/10.1016/j.ijengsci.2025.104457},
    url = {https://www.sciencedirect.com/science/article/pii/S0020722525002435},
    author = {R. Cavuoto and S. Scala and A. Cutolo and G. Mensitieri and M. Fraldi}
}

@article{Wang2017,
    doi = {10.1371/journal.pone.0188302},
    author = {Wang, Bo AND Tran, Huong AND Smith, Matthew A. AND Kostanyan, Tigran AND Schmitt, Samantha E. AND Bilonick, Richard A. AND Jan, Ning-Jiun AND Kagemann, Larry AND Tyler-Kabara, Elizabeth C. AND Ishikawa, Hiroshi AND Schuman, Joel S. AND Sigal, Ian A. AND Wollstein, Gadi},
    journal = {PLOS ONE},
    publisher = {Public Library of Science},
    title = {In-vivo effects of intraocular and intracranial pressures on the lamina cribrosa microstructure},
    year = {2017},
    month = {11},
    volume = {12},
    url = {https://doi.org/10.1371/journal.pone.0188302},
    pages = {1-16},
    number = {11}
}

@article{guidoboni2014,
  author  = {Guidoboni, G. and Harris, A. and Cassani, S. and Arciero, J. and Siesky, B. and Amireskandari, A. and Tobe, L. and Egan, P. and Januleviciene, I. and Park, J.},
  title   = {Intraocular Pressure, Blood Pressure, and Retinal Blood Flow Autoregulation: A Mathematical Model to Clarify Their Relationship and Clinical Relevance.},
  journal = {Investigative Ophthalmology \& Visual Science},
  volume  = {55},
  number  = {7},
  pages   = {4105--4118},
  year    = {2014},
  doi     = {10.1167/iovs.13-13611}
}

@article{Yan1998,
  author  = {Yan, D. B. and Flanagan, J. G. and Farra, T. and Trope, G. E. and Ethier, C. R.},
  title   = {Study of Regional Deformation of the Optic Nerve Head Using Scanning Laser Tomography},
  journal = {Current Eye Research},
  volume  = {17},
  number  = {9},
  pages   = {903--916},
  year    = {1998},
  doi     = {10.1076/ceyr.17.9.903.5143}
}

@misc{Okonkwo2026,
  author       = {Okonkwo, Ogugua N. and Tripathy, Koushik},
  title        = {Ocular Hypotony},
  year         = {2026},
  note         = {[Updated 2023 Aug 25]. In: StatPearls [Internet]},
  publisher    = {StatPearls Publishing},
  address      = {Treasure Island (FL)},
  url          = {https://www.ncbi.nlm.nih.gov/books/NBK582144/},
  urldate      = {2026-04-16}
}

@article{Girkin2024,
  author  = {Girkin, Christopher A. and Strickland, Ryan G. and Somerville, McKenna M. and Garner, Mary A. and Grossman, Gregory H. and Blake, Alan and Kumar, Nilesh and Ianov, Lara and Fazio, Massimo A. and Clark, Mark E. and Gross, Alecia K.},
  title   = {Acute Ocular Hypertension in the Living Human Eye: Model Description and Initial Cellular Responses to Elevated Intraocular Pressure},
  journal = {Vision Research},
  volume  = {223},
  pages   = {108465},
  year    = {2024},
  doi     = {10.1016/j.visres.2024.108465}
}

@article{Shen2025,
  author  = {Shen, Ru Yue and Zhang, Yuqiao and Chen, Li Jia and Cheung, Carol Y. and Liang, Yuanbo and Tham, Clement C. and Chan, Poemen P.},
  title   = {Ocular and Systemic Risk Factors and Biomarkers for Primary Glaucoma: An Umbrella Review of Systematic Reviews With Meta-Analyses},
  journal = {Investigative Ophthalmology \& Visual Science},
  volume  = {66},
  number  = {12},
  pages   = {35},
  year    = {2025},
  doi     = {10.1167/iovs.66.12.35}
}

@article{Jonas2013,
  author  = {Jonas, J. B. and Wang, N. and Yang, D. and Ritch, R.},
  title   = {Facts and Figures of Normal-Tension Glaucoma},
  journal = {Investigative Ophthalmology \& Visual Science},
  volume  = {54},
  number  = {13},
  pages   = {7931--7937},
  year    = {2013}
}

@article{Abe2015,
  author  = {Abe, R. Y. and Gracitelli, C. P. and Diniz-Filho, A. and Tatham, A. J. and Medeiros, F. A.},
  title   = {Lamina Cribrosa in Glaucoma: Diagnosis and Monitoring},
  journal = {Current Ophthalmology Reports},
  volume  = {3},
  number  = {2},
  pages   = {74--84},
  year    = {2015},
  doi     = {10.1007/s40135-015-0067-7}
}

@article{Ryu2024,
  author  = {Ryu, J. and Asaoka, R. and Nakakura, S. and Murata, H. and Nakaniida, Y. and Ishii, K. and Obana, A. and Kiuchi, Y.},
  title   = {Usefulness of Intraocular Pressure With the Ocular Response Analyzer to Predict Postoperative Hypotony Complications in Glaucoma},
  journal = {Translational Vision Science \& Technology},
  volume  = {13},
  number  = {10},
  pages   = {7},
  year    = {2024},
  doi     = {10.1167/tvst.13.10.7}
}

@article{hashin1965,
    author = {Hashin, Z.},
    title = {On elastic behaviour of fibre reinforced materials of arbitrary transverse phase geometry},
    journal = {Journal of the Mechanics and Physics of Solids},
    volume = {13},
    issue = {3},
    year = {1965},
    pages = {119-134}
}

@article{day1992,
    author = {A.R. Day and K.A. Snyder and E.J. Garboczi and M.F. Thorpe},
    title = {The elastic moduli of a sheet containing circular holes},
    journal = {Journal of the Mechanics and Physics of Solids},
    volume = {40},
    issue = {5},
    pages = {1031-1051},
    year = {1992}
}

@article{parnell2009,
    author = {Parnell WJ and Grimal Q},
    title = {The influence of mesoscale porosity on cortical bone anisotropy. Investigations via asymptotic homogenization.},
    journal = {J R Soc Interface},
    volume = {6},
    pages = {97-109},
    year = {2009}
}

@article{tan2018,
    author = {Tan, N.Y.Q. and Koh, V. and Girard, M.J.A. and Cheng, C.Y.},
    title = {Imaging of the lamina cribrosa and its role in glaucoma: A review},
    journal = {Clin Exp Ophthalmol},
    volume = {46},
    issue = {2},
    pages = {177-188},
    year = {2018}
}

\end{document}